\documentclass[12pt]{article}
\usepackage{amsmath}
\usepackage{graphicx,psfrag,epsf}
\usepackage{enumerate}
\usepackage{url}  
\usepackage{color} 

\usepackage{makecell}
\usepackage{multicol}
\usepackage{amssymb}
\usepackage{booktabs}
\usepackage[sort&compress,square,comma,authoryear]{natbib}
\usepackage{graphicx}
\usepackage{subfig}
\usepackage{tabularx}
\usepackage{algorithm}
\usepackage{algorithmic}
\usepackage{arydshln}
\usepackage{multirow}
\usepackage[nolists]{endfloat}

\setcitestyle{round}
\newcommand{\blind}{1}

\newcommand{\ra}[1]{\renewcommand{\arraystretch}{#1}}

\newcommand{\E}{\mathbb{E}}

\newcommand{\Prob}{\mathbb{P}}
\newcommand{\var}{\text{Var}}
\newcommand{\cov}{\text{Cov}}
\newcommand{\corr}{\text{Corr}}
\newcommand{\logit}{\text{logit}}
\newcommand{\expit}{\text{expit}}
\newcommand{\diag}{\text{diag}}
\newcommand{\ba}{\boldsymbol{\alpha}}
\newcommand{\bb}{\boldsymbol{\beta}}
\newcommand{\bg}{\boldsymbol{\gamma}}

\newcommand{\bk}{\boldsymbol{\kappa}}

\newcommand{\bpi}{\boldsymbol{\pi}}
\newcommand{\br}{\boldsymbol{\rho}}

\newcommand{\specialcell}[2][c]{%
	\begin{tabular}[#1]{@{}c@{}}#2\end{tabular}}
\allowdisplaybreaks
\begin{document}
\makeatletter
\def\adl@drawiv#1#2#3{%
	\hskip.5\tabcolsep
	\xleaders#3{#2.5\@tempdimb #1{1}#2.5\@tempdimb}%
	#2\z@ plus1fil minus1fil\relax
	\hskip.5\tabcolsep}
\newcommand{\cdashlinelr}[1]{%
	\noalign{\vskip\aboverulesep
		\global\let\@dashdrawstore\adl@draw
		\global\let\adl@draw\adl@drawiv}
	\cdashline{#1}
	\noalign{\global\let\adl@draw\@dashdrawstore
		\vskip\belowrulesep}}
\makeatother

\def\spacingset#1{\renewcommand{\baselinestretch}%
{#1}\small\normalsize} \spacingset{1}


\if1\blind
{
  \title{\bf A stochastic second-order generalized estimating equations approach for estimating intraclass correlation coefficient in the presence of informative missing data}
  \author{Tom Chen\\
    Department of Biostatistics, Harvard T.H.Chan School of Public Health\\
    and \\ 
	Eric J. Tchetgen Tchetgen \\ Department of Biostatistics, Harvard T.H.Chan School of Public Health \\ and Department of Epidemiology, Harvard T.H.Chan School of Public Health \\ and \\ 
	Rui Wang\thanks{rwang@hsph.harvard.edu. The authors gratefully acknowledge NIH grants T32ES007142 and R37AI51164}\hspace{.2cm}\\
	Department of Population Medicine, Harvard Medical School \\and Harvard Pilgrim Health Care Institute;\\Department of Biostatistics, Harvard T.H.Chan School of Public Health \\}
  \maketitle
} \fi

\if0\blind
{
  \bigskip
  \bigskip
  \bigskip
  \begin{center}
    {\LARGE\bf A stochastic second-order generalized estimating equations approach for estimating intraclass correlation coefficient in the presence of informative missing data}
\end{center}
  \medskip
} \fi

\begin{abstract}
Design and analysis of cluster randomized trials must take into account correlation among outcomes from the same clusters. When applying standard generalized estimating equations (GEE), the first-order (e.g. treatment) effects can be estimated consistently even with a misspecified correlation structure. In settings for which the correlation is of interest, one could estimate this quantity via second-order generalized estimating equations (GEE2). We build upon GEE2 in the setting of missing data, for which we incorporate a ``second-order" inverse-probability weighting (IPW) scheme and ``second-order" doubly robust (DR) estimating equations that guard against partial model misspecification. We highlight the need to model correlation among missing indicators in such settings. In addition, the computational difficulties in solving these second-order equations have motivated our development of more computationally efficient algorithms for solving GEE2, which alleviates reliance on parameter starting values and provides substantially faster and higher convergence rates than the more widely used deterministic root-solving methods.
\end{abstract}


\noindent
{\it Keywords:}  GEE, second-order, double robustness, algorithms
\vfill

\newpage
\spacingset{1.45} 
\section{Introduction}
\label{sec:intro}

	Cluster randomized trials (CRTs), in which individuals are randomly assigned to the intervention in groups, have been increasingly implemented to evaluate efficacy and effectiveness of various intervention programs. Design and analysis of CRTs must take into account possible correlation of outcomes within randomized units. The intraclass correlation coefficient (ICC) measures the degree to which individuals within a community are more similar to one another than to individuals in other communities and is crucial to accurately compute sample sizes needed to achieve a certain power level in a CRT.  The statistical power and required sample size for a CRT can change substantially depending on the ICC.  For example, in a matched-pair CRT with 15 pairs and a sample size of 300 within each cluster as in the Botswana Combination Prevention Project (BCPP) \citep{wang2014sample,gaolathe2016botswana}, the power to detect a 40\% reduction in 3-year cumulative incidence from 2.5\% to 1.5\% decreases from 80\% to 52\% as the ICC increases from 0.001 to 0.005. To achieve 80\% power with an ICC of 0.005, assuming all else being fixed, the number of clusters required is almost doubled (15 pairs to 27 pairs). When analyzing data from CRTs, a commonly used and robust approach is based on comparisons of a community-level measure of the end of interest. Tests constructed by giving equal weight to each cluster may not be fully efficient, especially when the sizes of clusters vary substantially. The optimal weights depend crucially on the ICC for both parametric test (e.g., t-test) \citep{hayes2009cluster} and nonparametric permutation tests \citep{braun2001optimal,wang2017use}. Despite its importance, obtaining reliable estimates of ICC remains a major problem in designing CRTs \citep{gail1992aspects,hayes1999simple,donner2000design,klar2001current}. Furthermore, ICC can vary considerably by intervention group and community characteristics (e.g., community size) \citep{wu2012comparison,crespi2009}.

	 In CRTs, interest often lies in estimating the causal effect of intervention on the cluster -- the difference between the outcome for the cluster when it receives intervention and the outcome when the cluster is untreated \citep{halloran1991,  carnegie2016estimation}. The generalized estimating equations (GEE) \citep{liang1986gee} approach provides an attractive option. This estimation procedure is semiparametric in that it does not require specification of a full likelihood, yet it can be made highly efficient by further specifying a working model for the conditional correlation structure (i.e. for ICC) of the correlated outcomes \citep{zeger1988models}.  Even with a misspecified ICC model, GEE still yields a consistent and asymptotically normal (CAN) estimator of the treatment effect, although estimators may no longer be efficient \citep{fitzmaurice1995caveat, wang2003misspecified}. As a result of this flexible feature, one typically estimates the ICC using moment estimators from the Pearson residuals \citep{mcdaniel2013fast}; when ICC is itself of primary interest, the method of moments approach can be inefficient and unreliable. This motivates us to consider more efficient estimators for the ICC which can be achieved via second-order generalized estimating equations (GEE2) \citep{zhao1990gee2, liang1992gee2}. 
	
	Several authors \citep{sutradhar2003, ziegler1998annotated} have noted of convergence problems regarding GEE2's, and we later demonstrate a much greater computational burden for GEE2 compared to GEE1. GEE2 are notoriously hard to solve due to the far larger stack of estimating equations for the association parameters, leading to excessive computing time for obtaining solutions to these equations. In our preliminary work, we found that when increasing the cluster sizes to 300 as in the BCPP, solving GEE2 becomes difficult due to both convergence issue and memory allocation issues. Furthermore, it is common to encounter missing outcomes in practice. When outcomes are assumed missing completely at random \citep{rubin1976} (MCAR; the outcomes are missing independently of both observed and unobserved data), GEE2 analysis performed on complete-case CRT data provides CAN estimators for the treatment and ICC parameters. In the case of missing at random (MAR; outcome missingness is independent of the unobserved variables conditional on the observed variables), GEE produces inconsistent estimates unless all factors contributing to the propensity of being missing are included in a correctly-specified outcome model. Currently, methods are available to account for a restricted missing at random mechanism (i.e. outcome missingness depends only on observed covariates but not on observed outcomes) in the GEE1 case for the estimation of marginal treatment effects through the use of inverse probability weighting (IPW) with augmentation of an outcome model (OM) \citep{prague2016accounting}. This augmented IPW approach falls under the general framework of doubly robust estimation \citep{robins1994regcoef, van2003unified, tsiatis2007semiparametric} and is doubly-robust (DR) in the sense that either the IPW model or OM need be correctly specified in order to produce consistent estimator of the treatment effect. However, how to extend the DR estimator in estimating the association parameters in the presence of missing data has not been investigated. Properly incorporating IPW for association parameters requires modeling the correlation among missingness indicators for correlated units within a cluster, a potential complication which to the best of our knowledge has previously not been considered in the literature on semiparametric methods for missing clustered data. \cite{robins1995analysis} modeled the joint missingness process in the context of longitudinal data. In the context of CRTs, there is no natural ordering of the outcomes within a community and the missingness pattern is non-monotone, making the problem much more intractable \citep{tsiatis2007semiparametric}.

    In this paper, we investigate the use of IPW in GEE2s (IPW-GEE2) to account for outcome-missing data. If the model for the missingness mechanism is estimated consistently, the first- and second-order IPW provide CAN estimators of both the mean and high-order association effects by re-weighting complete cases according to the probability of being observed \citep{liang1986gee, robins1994regcoef}. To guard against misspecification of the IPW model, we further propose a doubly-robust GEE2 estimator (DR-GEE2), which, similar to  \cite{prague2016accounting}, produces consistent estimators for the mean and association parameters if either the IPW model or OM is correctly specified.
    
   Another purpose of this paper is to develop stochastic methods to alleviate the computational challenges associated with solving GEE2. These stochastic algorithms involve running Fisher scoring on a different subset of the data at each iteration, in the spirit of minibatch stochastic gradient descent (mbSGD) and the more general class of Robbins-Monro (RM) algorithms.  Under mild regularity conditions \citep{blum1954}, the algorithm almost surely converges to the same solution as if we performed standard Fisher scoring on GEE2. However, in the setting of correlated data subject to informative missingness, one cannot naively cycle through the subset of equations because some equations are given more importance than others, depending on the IPW and cluster characteristics. This unique combination not only suggests, but requires the use of informative sampling schemes in properly cycling through the data.

	In Section \ref{sec:methods}, we introduce GEE2 in the absence of missing data, and subsequently consider IPW-GEE2 and DR-GEE2 to account for missing outcome data. Definitions of marginalized ICC, model parametrization for GEE2, and joint models for the missing data process are discussed in this Section. In Section \ref{sec:stochastic}, we introduce the RM algorithm and expand on the stochastic paradigm to model fitting, and adapt this approach to fitting GEE2, which we coin as stochastic GEE2. Issues such as computational complexity, efficient implementation, and parallelization  as a further mechanism in reducing computing time and computing error are explored here. We evaluate the performance of the proposed estimators and the proposed computational algorithms with simulations in Section \ref{sec:simulation} and apply the new estimators and algorithms to analyze the Bangladeshi sanitation data in Section \ref{sec:sanitation}. We end with a discussion in Section \ref{sec:discusion}. Proofs are relegated to the Appendix.

\section{Methods}
\label{sec:methods}
\subsection{Notation and Models} \label{sec:notmodel}
	Henceforth, we work with binary outcomes $Y_{ij} \in  \{0, 1\}$ for subject $j = 1, \cdots, n_i$ in cluster $i = 1, \cdots, I$; the framework is readily generalizable to continuous outcomes. Let $A_i \in \{0, 1\}$ denote the treatment randomized at the cluster level with $\Prob(A_i = 1) = p_A$; $\textbf{Z}_i \in \mathbb{R}^{q}$ and $\textbf{X}_{ij} \in \mathbb{R}^m$ as the baseline cluster- and subject-level covariates, respectively; and $\textbf{X}_i = \{\textbf{X}_{ij}\}_{j=1}^{n_i}$. We denote $P(\cdot)$ as the probability measure associated with the argument i.e. $P(a), P(\textbf{z}, \textbf{x})$. Let $\pi_{ij} = \E[Y_{ij}|A_i, \textbf{Z}_i, \textbf{X}_i]$ denote the conditional mean outcome and 
	\begin{align*}
	\rho_{ijj'} = \corr(Y_{ij}, Y_{ij'}|A_i,\textbf{Z}_i,\textbf{X}_i) \overset{\text{def}}{=} \cov(Y_{ij}, Y_{ij'}|A_i,\textbf{Z}_i,\textbf{X}_i)\bigg/\sqrt{\var(Y_{ij}|A_i,\textbf{Z}_i,\textbf{X}_i)\var(Y_{ij'}|A_i,\textbf{Z}_i,\textbf{X}_i)}
	\end{align*}
	denote the conditional ICC. The quantities of interest are $\pi_i^* = \E[Y_{ij}|A_i]$ and $\rho_i^* = \corr(Y_{ij}, Y_{ij'}|A_i)$, which are the treatment-specific mean outcome and ICC. It is clear that $\pi_i^*$ is a marginalization of $\pi_{ij}$ in the sense that $\pi_{i}^* = \E[\pi_{ij}|A_i] = \int \pi_{ij} dP(\textbf{z}_i, \textbf{x}_i)$. But,	$\rho_{i}^* \neq \E[\rho_{ijj'}|A_i]$ in general. Indeed, it is easy to confirm that $\rho_i^* = \E[\rho_{ijj'}^\dagger|A_i]$, where
		\begin{equation}\label{rhodag}
		\begin{aligned}
		\rho_{ijj'}^\dagger \overset{\text{def}}{=}\E\left[\frac{(Y_{ij}- \pi_{i}^*)(Y_{ij'}- \pi_{i}^*)}{\pi_{i}^*(1-\pi_{i}^*)}\bigg|A_i,\textbf{Z}_i, \textbf{X}_i\right] = \frac{(\pi_{ij}-\pi_{i}^*)(\pi_{ij'}-\pi_{i}^*) + \rho_{ijj'} \sqrt{\mathcal{V}_{ijj'}}}{\pi_{i}^*(1-\pi_{i}^*)}
		\end{aligned}
		\end{equation}
	where $\mathcal{V}_{ijj'} = \pi_{ij}(1-\pi_{ij})\pi_{ij'}(1-\pi_{ij'})$. 

	Let $\widehat{\pi}_{ij}$ be an estimator of $\pi_{ij}$, converging to the limit $\overline{\pi}_{ij}$, which may or may not equal the true $\pi_{ij}$. Likewise, define $\widehat{\rho}_{ijj'}$ and $\overline{\rho}_{ijj'}$. Standard models for $\widehat{\pi}_{ij}$ include logistic or probit regression, while a model for $\widehat{\rho}_{ijj'}$ would be a generalized linear model with link function $g(x) = \text{atanh}(x)$, the Fisher $z$-transform. The Fisher $z$-transform is commonly used as a variance-stabilizing transformation for the sample correlation coefficient, but we apply it here to map the $[-1, 1]$ support of $\rho_i^*$ onto $\mathbb{R}$.
	
	Similarly, let $\widehat{\pi}_i^*$ and $\widehat{\rho}_i^*$ be estimators for $\pi_i^*$ and $\rho_i^*$ with limits $\overline{\pi}_i^*$ and $\overline{\rho}_i^*$, respectively. For example, inference for the effect of $A_i$ can be estimated under the model
	\begin{equation} \label{marginal}
	\begin{aligned}
	\logit(\pi_{i}^*(\bb_{Y}^*; A_i)) &=  \beta_{0Y}^* + \beta_{AY}^*A_i\\
	\text{atanh}(\rho_i^*(\ba_{Y}^*; A_i)) &=  \alpha_{0Y}^* + \alpha_{AY}^*A_i
	\end{aligned}
	\end{equation}
	to produce estimators $(\widehat{\bb}_{Y}^*, \widehat{\ba}_{Y}^*)$. Eq \ref{marginal} will be referred to as the canonical treatment model (TM). In the absence of missing data, and since $A_i$ is binary, the canonical TM is guaranteed to yield consistent $\overline{\pi}_i^* = \pi_i^*$ and $\overline{\rho}_i^* = \rho_i^*$. In the standard GEE2 framework, we would estimate $(\widehat{\bb}^*_Y, \widehat{\ba}^*_Y)$  as the solution to the equations
	 \begin{equation}
	\begin{aligned}\label{gee2}
	\textbf{0} &= \sum_{i=1}^{I} D_{i}^\intercal V_{i}^{-1}E_{i} \overset{\text{def}}{=} \sum_{i=1}^{I} \textbf{S}_i^Y(A_i, \bb_Y^*, \ba_Y^*)
	\end{aligned}
	\end{equation}
	where 
	\begin{align*}
	D_{i} &= \frac{\partial(\bpi_i^*(\bb^*_Y; A_i), \br_i^*(\ba^*_Y; A_i))}{\partial  (\bb_Y^*, \ba_Y^*)^\intercal} \qquad  V_{i}= \cov\begin{pmatrix}  \textbf{Y}_i \\ \mathcal{E}(\textbf{Y}_i) \end{pmatrix} \qquad E_{i} = \begin{pmatrix} \textbf{Y}_i - \bpi_i^*(\bb^*_Y)\\ \mathcal{E}(\textbf{Y}_i) - \br_i^*(\ba^*_Y)\end{pmatrix}
	\end{align*}
	and
	\begin{align*}
	\mathcal{E}(\textbf{Y}_i) &= \left[\frac{(Y_{ij}- \pi_{i}^*)(Y_{ij'}- \pi_{i}^*)}{\pi_{i}^*(1-\pi_{i}^*)}\right]_{j < j'} 
	\end{align*}
	Note that the working covariance matrix $V_i$ need not be correctly specified to produce consistent estimates, but doing so may lead to improved efficiency. We discuss forms of $V_i$ in Section \ref{sec:stochastic}. The expression above involving the standardized residuals $\mathcal{E}(\textbf{Y}_i)$ is one particular parametrization of GEE2 \citep{ziegler2000familial}, but we note there are others \citep{liang1992gee2, zhao1990gee2}. We pick the above parametrization because it specifically targets estimating the treatment-specific ICC $\rho_{i}^*$ instead of, say, the cross moments or covariances as in the other parametrizations. The focus of this paper is on making valid inferences about the treatment-specific mean and ICC, as quantified by $({\bb}^*_Y, {\ba}^*_Y)$, in the presence of missing data.\\
	
	\subsection{IPW-GEE2}\label{sec:ipwgee2}
	Accounting for missing outcome data in CRTs is challenging under the missing at random (MAR) assumption because there is no natural ordering of the outcomes within a cluster and the missingness can not be considered as monotone. We consider a submodel of MAR, restricted MAR (rMAR) as in \cite{prague2016accounting}. If $R_{ij}$ is the missingness indicator for $Y_{ij}$ with $R_{ij} = 0$ indicating $Y_{ij}$ is missing, then rMAR is equivalent to $\Prob(R_{ij} = 1|\textbf{Y}_i, A_i, \textbf{Z}_i, \textbf{X}_i) = \Prob(R_{ij} = 1|A_i, \textbf{Z}_i, \textbf{X}_i)$. To continue with valid inference, we assume that $\Prob(R_{ij} = 1|A_i, \textbf{Z}_i, \textbf{X}_i) > 0$, commonly known as the positivity assumption (PO). We propose the inverse-probability weighting second-order generalized estimating equations (IPW-GEE2) as
	\begin{equation} \label{ipwgee2}
	\begin{aligned}
	\textbf{0} &=  \sum_{i=1}^{I} D_{i}^\intercal V_{i}^{-1}W_{i}^RE_{i} \overset{\text{def}}{=} \sum_{i=1}^{I} \Phi_i^Y(A_i, \bb_Y^*, \ba_Y^*, \bb_R, \ba_R) \\
	\textbf{0} &= \sum_{i=1}^{I} \textbf{S}_i^R(A_i,\textbf{Z}_i, \textbf{X}_i, \bb_R, \ba_R)
	\end{aligned}
	\end{equation}
	where we have incorporated the following inverse-probability weighting matrix:
	\begin{align*}
	W_{i}^R = \diag\left(\underbrace{\frac{R_{i1}}{\overline{\pi}^R_{i1}(\bb_R)},\cdots,\frac{R_{in_i}}{\overline{\pi}^R_{in_i}(\bb_R)}}_{\text{IPW1}}, \underbrace{\frac{R_{i1}R_{i2}}{\overline{\eta}^R_{i12}(\bb_R,\ba_R)}, \cdots, \frac{R_{i(n_i-1)}R_{in_i}}{\overline{\eta}^R_{i(n_i-1)n_i}(\bb_R,\ba_R)}}_{\text{IPW2}}\right)
	\end{align*}
	$\textbf{S}_i^R$ is structurally the same as Eq \ref{gee2}, except with a full model for $\textbf{R}_i$ instead of a treatment-specific model for $\textbf{Y}_i$. Here, $(\bb_R, \ba_R)$ are nuisance parameters that must be estimated, but of no interest for inference. Within the IPW matrix, $\overline{\pi}^R_{ij}(\bb_R)$ is a model (parametrized by $\bb_R$) for $ \pi^R_{ij} = \Prob(R_{ij} = 1|A_i, \textbf{Z}_i, \textbf{X}_i)$ and $\overline{\eta}_{ijj'}(\bb_R, \ba_R)$ is a model (parametrized by $\bb_R, \ba_R$) for $\eta^R_{ijj'} = \Prob(R_{ij} = R_{ij'} = 1|A_i, \textbf{Z}_i, \textbf{X}_i)$; we shall refer to them as the first-order and second-order propensity scores (PS1 \& PS2), respectively. Since $\eta_{ijj'}^R$ is a function of $\pi_{ij}^R, \pi_{ij'}^R, \rho_{ijj'}^R$, it suffices to fit a model for $\rho_{ijj'}^R$. $W_{i}^R$ itself is the inverse-probability weighting (IPW) matrix, which can be decomposed into IPW1 and IPW2 portions. We refer to the first equation of Eqs \ref{ipwgee2}  as the treatment model estimating equation (TMEE) portion, while the second equation of Eqs \ref{ipwgee2}, which produce estimators $\widehat{\pi}_{ij}^R$ (converging to $\overline{\pi}_{ij}^R$) and $\widehat{\rho}_{ijj'}^R$ (converging to $\overline{\rho}_{ijj'}^R$), as the propensity score estimating equation (PSEE) portion. 

	IPW-GEE1 been explored before in \cite{prague2016accounting}. The IPW2 portion is derived by considering that the $(j, j')$th element of $\mathcal{E}(\textbf{Y}_i)$ is missing when either $Y_{ij}$ or $Y_{ij'}$ is missing; this is exactly represented by the product of their missingness indicators, $R_{ij}R_{ij'}$, for which we would then need to model $\eta_{ijj'}^R(\bb_R, \ba_R)$. To the best of our knowledge, this is the first instance in which a model is required for the joint missingness indicator $R_{ij}R_{ij'}$ in the context of clustered data. Not properly accounting for the correlation among missingness indicators will in general lead to biased estimates for the association parameters. Unlike the treatment model, the PS can possibly be misspecified; if so, then estimators $(\widehat{\bb}_Y^*, \widehat{\ba}_Y^*)$ may not be consistent. 

	\subsection{DR-GEE2}\label{sec:drgee2}
	The augmented GEE (AUG) methods, which adds a term to the standard GEE that relates the outcome to covariates and treatment, have been proposed to improve estimation efficiency by leveraging baseline covariates in the setting of CRTs \citep{stephens2012}. \cite{prague2016accounting} proposed a doubly robust estimator based on augmentation for estimating the marginal treatment effect in CRTs when data are rMAR to guard against misspecification of either the OM and PSM. Here we extend to the GEE2 framework, which we call DR-GEE2:
	\begin{equation} 
	\begin{aligned}\label{augipwgee2}
	\textbf{0} &=\sum_{i=1}^{I} [D_i^\intercal V_i^{-1}W_{i}^RE_i' +\zeta_i]\overset{\text{def}}{=} \sum_{i=1}^{I} \widetilde{\Phi}_i^Y(\textbf{Z}_i^*, \textbf{X}_i,\textbf{R}_i, \bb_Y^*, \ba_Y^*,\bb_R, \ba_R, \bb_Y, \ba_Y) \\
	\textbf{0} &= \sum_{i=1}^{I} \textbf{S}_i^R(\textbf{Z}_i^*, \textbf{X}_i, \bb_R, \ba_R) \\
	\textbf{0} &= \sum_{i=1}^{I} \textbf{S}_i^Y(\textbf{Z}_i^*, \textbf{X}_i, \bb_Y, \ba_Y)
	\end{aligned}
	\end{equation}
	where
	\begin{align*}
	E_i' &= \begin{pmatrix} \textbf{Y}_i - \overline{\bpi}_i(\bb_Y) \\ \mathcal{E}(\textbf{Y}_i) - \overline{\br}_i^\dagger(\ba_Y)\end{pmatrix}, 	\qquad E_i'' = \begin{pmatrix} \overline{\bpi}_i(\bb_Y) - \bpi_i^*(\bb_Y^*) \\ \overline{\br}_i^\dagger(\ba_Y) - \br_i^*(\ba_Y^*)\end{pmatrix} \\
	\zeta_i &=  \sum_{a=0}^{1}p^a_A(1-p_A)^{1-a} D_i^\intercal (A = a)V_i^{-1}E_i''(A = a)  
	\end{align*}
	where $\overline{\pi}_{ij}$ is a model for $\pi_{ij}$ and
	\begin{equation*}
	\begin{aligned}
	\overline{\rho}_{ijj'}^\dagger = \frac{(\overline{\pi}_{ij}-\overline{\pi}_{i}^*)(\overline{\pi}_{ij'}-\overline{\pi}_{i}^*) + \overline{\rho}_{ijj'} \sqrt{\overline{\mathcal{V}}_{ijj'}}}{\overline{\pi}_{i}^*(1-\overline{\pi}_{i}^*)}
	\end{aligned}
	\end{equation*}
	akin to Eq \ref{rhodag}, with models replacing each population quantity. The third set of equations in Eq \ref{augipwgee2}, which we refer to as the outcome model estimating equations (OMEE), fits $\widehat{\pi}_{ij}$ (converging to $\overline{\pi}_{ij}$) and $\widehat{\rho}_{ijj'}$ (converging to $\overline{\rho}_{ijj'}$), collectively known as the outcome models. If the OM are correctly specified, then under the rMAR assumption, $(\bb_Y, \ba_Y)$ can be consistently estimated based on the complete-case data. The DR estimator is doubly robust in the sense that it is CAN under correct specification of either the OM [i.e. $\overline{\pi}_{ij} = \pi_{ij}$ and $\overline{\rho}_{ijj'} = \rho_{ijj'}$] or PS [i.e. $\overline{\pi}_{ij}^R = \pi_{ij}^R$ and $\overline{\rho}_{ijj'}^R = \rho_{ijj'}^R$] (see proof in Appendix \ref{appendixDR}). 
	\subsection{Inference}\label{sec:inference}
	Variance of $(\widehat{\bb}_Y^*, \widehat{\ba}_Y^*)$ is estimated by the sandwich estimator. Denote $\bk = (\bb^*_Y, \ba^*_Y, \bb_R, \ba_R, \bb_Y, \ba_Y)$ and
		\begin{align*}
		\Psi(\bk) = \begin{pmatrix}
		\widetilde{\Phi}_i^Y(A_i,\textbf{Z}_i, \textbf{X}_i,\textbf{R}_i, \bb_Y^*, \ba_Y^*,\bb_R, \ba_R, \bb_Y, \ba_Y) \\
		\textbf{S}_i^R(A_i,\textbf{Z}_i, \textbf{X}_i, \bb_R, \ba_R) \\
		\textbf{S}_i^Y(A_i,\textbf{Z}_i, \textbf{X}_i, \bb_Y, \ba_Y)
		\end{pmatrix}
		\end{align*}
	A standard Taylor expansion paired with Slutsky's theorem and the central limit theorem provide the DR-GEE2 sandwich estimator adjusted for estimation of nuisance parameters in the OM and PS: $\var(\widehat{\bk}) = \Gamma^{-1} \Delta (\Gamma^{-1})^\intercal$, where $\Delta(\bk) = \E\left[\Psi(\bk) \Psi(\bk)^\intercal\right]$ and $\Gamma(\bk) = \E\left[\partial\Psi(\bk)/\partial \bk^\intercal\right]$, from which we can extract components corresponding to just $(\widehat{\bb}^*_Y, \widehat{\ba}^*_Y)$.  An estimator $\widehat{\var}(\widehat{\bk})$ can be obtained by replacing $\Delta$ with $\widehat{\Delta} = \frac{1}{I}\sum_{i=1}^{I}\widehat{\Psi}(\widehat{\bk})\widehat{\Psi}(\widehat{\bk})^\intercal$ and $\Gamma$ with $\widehat{\Gamma} = \frac{1}{I}\sum_{i=1}^{I}\partial \widehat{\Psi}(\widehat{\bk})/\partial \bk$.

	\section{A stochastic algorithm for solving GEE2's} \label{sec:stochastic}
	In this section, we make the following assumption regarding the working covariance matrix for GEE2, similar to \cite{yan2004} in their R package \texttt{geepack}:
	 $\cov(\textbf{Y}_i, \mathcal{E}(\textbf{Y}_i)) = \textbf{0}_{n_i \times \binom{n_i}{2}}$ and $\var(\mathcal{E}(\textbf{Y}_i)) = \mathcal{I}_{\binom{n_i}{2}}$, and similarly for the working correlation structure on the PSEE and OMEE. That is, we are imposing a working correlation structure in our GEE2 where the off-diagonal blocks are all zeros, and the lower-right block corresponding to variance-covariance components of $\mathcal{E}(\textbf{Y}_i)$ is just the identity matrix. This latter assumption is commonly done in practice due to the difficulty in specifying models for higher moments. We include the treatment-specific ICC estimates from the GEE2 embedded within the working correlation structure $\var(\textbf{Y}_i)$ of the GEE1 portion.  Correct specification of the working correlation structure for GEE in the absence of missing data is theoretically optimal and have been demonstrated in simulations to have vast efficiency gains \citep{fitzmaurice1995caveat}, while cases have also been noted where the use of independence correlation structure is just as efficient \citep{zeger1988regression, mcdonald1993estimating}. 
	 
	These additional assumptions allow us to separate our IPW/DR-GEE2 equations for $Y_{ij}$ into two portions:
	\begin{equation}
	\begin{aligned}
	\textbf{0} &= \sum_{i=1}^{I}G_{\bb i} \overset{\text{def}}{=} G_{\bb} \qquad \qquad \text{GEE1 portion}\\
	\textbf{0} &= \sum_{i=1}^{I}G_{\ba i} \overset{\text{def}}{=} G_{\ba} \qquad  \qquad\text{GEE2 portion}
	\end{aligned}
	\end{equation}
	where gradient $G_{\bb i}$ equals the GEE1 portion of either $\Phi_i^Y$ in Eq \ref{ipwgee2} or $\widetilde{\Phi}_i^Y$ in Eq \ref{augipwgee2}, and similarly for $G_{\ba i}$.
	Define $H_{\bb} = -\E\left[\frac{d}{d\bb^\intercal} G_{\bb}\right]$ and $H_{\ba} = -\E\left[\frac{d}{d\ba^\intercal} G_{\ba}\right]$ as the expected Fisher information (negative Hessian) of the $\bb, \ba$ components. Then the Fisher scoring (Newton-Raphson) iterations to solve the IPW-GEE2 take the following form:
	\begin{align*}
	\bb_{\omega+1} &= \bb_{\omega} + H^{-1}_{\bb(\omega)}G_{\bb(\omega)}\\
	\ba_{\omega+1} &= \ba_{\omega} + H^{-1}_{\ba(\omega)}G_{\ba(\omega)}
	\end{align*}
	Each iteration of the GEE1 portion involves vectors and square matrices of dimension $n_i$ and $n_i \times n_i$, respectively. The GEE2 portion involves dimension $\binom{n_i}{2}$ and $\binom{n_i}{2} \times \binom{n_i}{2}$ vectors/matrices, which do not scale well and lead to the aforementioned convergence rate and convergence time problems. Our solution is to refine Fisher scoring with the Robbins-Monro (RM) algorithm \citep{robbins1951}.
	\subsection{Background: Robbins-Monro Algorithm} \label{sec:robbins}
	The Robbins-Monro (RM) algorithm \citep{robbins1951} states that, in solving for the zero $\theta_0$ in the equation $\psi(\theta) = 0$, if we instead have the random variable $\phi(\theta)$ such that $\E[\phi(\theta)] = \psi(\theta)$, then we may iterate
		\begin{align*}
		\theta_{\omega+1} = \theta_{\omega} - \gamma_\omega \phi(\theta_\omega)
		\end{align*}
	where learning rates $\gamma_\omega > 0$ satisfy $\sum_{\omega} \frac{1}{\gamma_\omega} = \infty$ and $\quad \sum_{\omega} \frac{1}{\gamma_\omega^2} < \infty$. Given these previous conditions, and a few other mild regularity conditions (collectively known as the Robbins-Monro conditions), we have that $\theta_{\omega} \rightarrow \theta_0$ in $L^2$-mean. \cite{blum1954} provides a proof that $\theta_{\omega} \rightarrow \theta_0$ almost surely. The RM algorithm is useful whenever we can find such a $\phi$ which is also significantly faster to compute than $\psi$. For example, consider the general $M$-estimation problem (for which GEE is a special case) and suppose our estimating equation takes the form $\psi(\theta) = \sum_{i=1}^{I}\psi_i(\theta)$. It is easy to confirm that
		\begin{align*}
		\phi(\theta) = \sum_{i \in s} \frac{\psi_i(\theta)}{p_i}
		\end{align*}
	satisfies $\E[\phi(\theta)] = \psi(\theta)$, where $s$ is a randomly chosen subset of $U = \{1, \cdots, I\}$ according to some sampling design $\mathbb{D}$ with $p_i = \Prob(i \in s)$. Here, instead of performing $I$ function evaluations, we only need to perform $|s|$ evaluations. If we take $\mathbb{D}$ to be a simple random sample without replacement (SRSWOR) of size $\upsilon$, this reduces to minibatch stochastic gradient descent (mbSGD) (see \cite{clemencon2015} for general sampling schemes). 
	\subsection{SGEE2}\label{sec:sgee2}
	In CRTs such as the Botswana Combination Prevention Project (BCPP) \citep{gaolathe2016botswana}, researchers are often faced with few clusters and large cluster sizes. Hence, the design of the proposed class of stochastic GEE2 (SGEE2) algorithm differs from the standard mbSGD in that we are improving iteration speed not through evaluating fewer of the functional summands $\{\psi_i\}_{i=1}^{I}$ (i.e. evaluating fewer clusters), but rather evaluating an unbiased and computational-easier estimate of each summand $\psi_i$ (done through sampling a subset of individuals per cluster). More intuitively, mbSGD is akin to cluster sampling, while SGEE2 is akin to stratified sampling.
	
	Another improvement of SGEE2 over the mbSGD framework is the inclusion of the Hessian. Much of the literature derived from the Robbins-Monro framework does not incorporate the Hessian matrix into the iterations, instead relying on adaptive gradients and adaptive learning rates \citep{nesterov1983, duchi2011, zeiler2012}. Traditionally, Hessians are omitted because they are hard to compute \citep{bottou2012}. The Hessians are simply the negative Fisher information, which in the GEE2 framework, is straightforward to calculate. We exploit this closed-form to arrive at an unbiased and computationally-easier estimate of the observed Hessians. Since we are estimating the Hessians as well, our SGEE2 algorithms also fall under the class of quasi-Newton or variable metric methods \citep{luksan2000}. 
	
	Even for simple functions, Fisher scoring / Newton-Raphson are known for divergence issues related to stationary points; that is, on the iteration trail to the solution of the gradient / score equations, there are evaluation points for which the Hessians / observed information are nearly zero. One way to overcome this barrier is by trying different initial values that avoid these stationary values. This technique is more formally known as multistart search \citep{ugray2007} and attempts to scatter starting points in hopes that a few are within the set of points which always converge to a solution, known as basins of attraction from the numerical analysis literature. In deterministic Fisher scoring, if one is within a basin of attraction, any future iteration point will also be within a basin of attraction by definition; the inverse is also true. SGEE2 naturally solves this issue because, even if one were not within a basin of attraction, the stochastic nature of the algorithm makes it very likely to ``jump" off the path of divergence back en route to a solution. This is a double-edged sword, because it may also be possible to be jerked off the path of convergence. This is mostly not an issue, because in practice the basins of attractions are often far larger than the basins of repellents, and our simulation study in Section \ref{algchar} confirms this.
	\subsection{S-IPW-GEE2} \label{sec:sipwgee2}
	The Fisher scoring for IPW-GEE2 equations have gradients and negative Hessians of the form
		\begin{equation} \label{SIPWGEE2}
		\begin{aligned}
		H_{\bb(\omega)} &= \sum_{i=1}^{I}D^\intercal_{\beta i(\omega)}V_{\beta i (\omega)}^{-1}W_{\beta i (\omega)}^R D_{\beta i(\omega)}, \qquad G_{\bb(\omega)} = \sum_{i=1}^{I}D^\intercal_{\beta i(\omega)}V_{\beta i (\omega)}^{-1}W_{\beta i (\omega)}^R E_{\beta i(\omega)}\\
		H_{\ba(\omega)} &= \sum_{i=1}^{I}D^\intercal_{\alpha i(\omega)}W_{\alpha i (\omega)}^R D_{\alpha i(\omega)}, \qquad \qquad G_{\ba(\omega)} = \sum_{i=1}^{I}D^\intercal_{\alpha i(\omega)}W_{\alpha i (\omega)}^R E_{\alpha i(\omega)}
		\end{aligned}
		\end{equation}
	For what we define as the standard S-IPW-GEE2, we take our universe $U^\text{obs} = (U_1^\text{obs}, \cdots, U_I^{\text{obs}})$, where each $U_i^\text{obs}$ correspond to the indices of the observed outcomes in cluster $i$, and let $m_i = |U_i^\text{obs}|$ be the number of non-missing observations per cluster. At each iteration $\omega$, sample $s_i \sim \text{SRSWOR}(U_i^\text{obs}, \upsilon_i)$, and concatenate $s = (s_1, \cdots, s_I)$. That is, each cluster sample $s_i$ is a simple random sample without replacement of $\upsilon_i$ indices of the nonmissing data. The default context chooses $\upsilon_i = \lceil \pi_S |U_i^\text{obs}|\rceil$ for some sampling proportion $\pi_S \in (0, 1)$. Notationally, we can treat $s$ as our observed sample, in which case defining stochastic versions $\widetilde{H}_{\beta i (\omega)}$, $\widetilde{G}_{\beta i (\omega)},$ $\widetilde{H}_{\alpha i (\omega)}$, and $\widetilde{G}_{\alpha i (\omega)}$ simply requires modifying the IPW matrices in the full Fisher scoring with the induced missingness from subsampling, resulting with $\widetilde{W}_{\beta i (\omega)}^R = \frac{m_i}{\upsilon_i}W_{\beta i (\omega)}^R [s_i]$ and $\widetilde{W}_{\alpha i (\omega)}^R = \frac{m_i(m_i-1)}{\upsilon_i(\upsilon_i-1)}W_{\alpha i (\omega)}^R[(s_i)_2]$, where $[s_i]$ is a 0--1 diagonal matrix indicating if observation $j$ is included in subsample $s_i$, and similarly defined with two-way combinations for $[(s_i)_2]$. It is easy to verify that 
	\begin{equation}\label{unbiasedgrad}
	\begin{aligned}
	\E[\widetilde{H}_{\beta (\omega)}|\mathcal{D}] &= \widehat{H}_{\beta (\omega)}, \qquad
	\E[\widetilde{G}_{\beta (\omega)}|\mathcal{D}] = \widehat{G}_{\beta (\omega)} \\
	\E[\widetilde{H}_{\alpha(\omega)}|\mathcal{D}] &= \widehat{H}_{\alpha (\omega)},\qquad
	\E[\widetilde{G}_{\alpha (\omega)}|\mathcal{D}] = \widehat{G}_{\alpha (\omega)} 
	\end{aligned}
	\end{equation}
	where $\mathcal{D}$ is the observed data and the expectation is taken with respect to the conditional law $P(s|\mathcal{D})$. The expressions in Eqs \ref{unbiasedgrad} are simply marginalizing out the induced randomness from choosing our subset $s$ of our given data. Hence, by the RM conditions, we have that S-IPW-GEE2 produces estimates $(\widetilde{\bb}, \widetilde{\ba}) \rightarrow (\widehat{\bb}, \widehat{\ba})$ almost surely with respect to the conditional law $P(s|\mathcal{D})$. Furthermore, the stochastic Hessians leverage information about the curvature of the objective function, hence providing faster convergence as well. We present the full details in pseudocode of S-IPW-GEE2 in Algorithm \ref{alg:IPWSGEE2} in Appendix \ref{appendixalgo}.
	\subsection{S-DR-GEE2} \label{sec:sdrgee2}
	The gradients and negative Hessians under DR-GEE2 are
	\begin{equation}\label{SDRGEE2}
	\begin{aligned}
	H_{\bb(\omega)} &= \sum_{i=1}^{I}\sum_{a=0}^{1}p^a(1-p)^{1-a}D^\intercal_{\beta i(\omega)}(A=a)V_{\beta i (\omega)}^{-1}D_{\beta i(\omega)}(A=a)\\
	G_{\bb(\omega)} &= \sum_{i=1}^{I}[D^\intercal_{\beta i(\omega)}V_{\beta i (\omega)}^{-1}W_{\beta i (\omega)}^RE_{\beta i(\omega)}' + \zeta_{\beta i(\omega)}]\\
	H_{\ba(\omega)} &= \sum_{i=1}^{I}\sum_{a=0}^{1}p^a(1-p)^{1-a}D^\intercal_{\alpha i(\omega)}(A=a)D_{\alpha i(\omega)}(A=a)\\
	G_{\ba(\omega)} &= \sum_{i=1}^{I}[D^\intercal_{\alpha i(\omega)}W_{\alpha i (\omega)}^RE_{\alpha i(\omega)}' + \zeta_{\alpha i(\omega)}]
	\end{aligned}
	\end{equation}
	The expressions are more complex than those from IPW-SGEE2 due to the addition of the augmentation term $\zeta_{\cdot i(\omega)}$. Structurally speaking, the PS term $E_i'$ comprises of the true data $Y_{ij}$ that can be missing, while the OM term $E_i''$ comprises of OM predictions that are never missing. Hence, in the construction of the S-DR-GEE2 algorithm, using the same subsample $s_i$ of indices of $E_i'$ for the indices of $E_i''$ would result in a biased estimator of $\zeta_{\cdot i(\omega)}$. Specifically, consider the following candidates for stochastic versions of $\zeta_{\cdot \beta(\omega)}$:
	\begin{align*}
	\zeta_{\cdot \beta(\omega)}^{(1)} &= \sum_{a=0}^{1}p_A^a(1-p_A)^{1-a} D_{\beta i(\omega)}^\intercal (A = a)V_{\beta i(\omega)}^{-1}\widetilde{W}_{\beta i(\omega)}^{R}E_{\beta i(\omega)}''(A = a) \\
	\zeta_{\cdot \beta(\omega)}^{(2)} &= \sum_{a=0}^{1}p_A^a(1-p_A)^{1-a} D_{\beta i(\omega)}^\intercal (A = a)V_{\beta i(\omega)}^{-1}\frac{m_i}{\upsilon_i}[s_i]E_{\beta i(\omega)}''(A = a) \\
	\zeta_{\cdot \beta(\omega)}^{(3)} &= \sum_{a=0}^{1}p_A^a(1-p_A)^{1-a} D_{\beta i(\omega)}^\intercal (A = a)V_{\beta i(\omega)}^{-1}\widetilde{W}_{\cdot i (\omega)}^{R'}E_{\beta i(\omega)}''(A = a)
	\end{align*}
	where $\widetilde{W}_{\cdot i (\omega)}^{R'} = \frac{m_i}{\upsilon_i'}[s_i']$ and $s_i' \subseteq \{1, \cdots, n_i\}$ denotes an independent sample of $\upsilon_i'$ indices for the entire cluster, not just the observed $U_i^\text{obs}$. In general, $\E[\zeta_{\cdot \beta(\omega)}^{(1)}|\mathcal{D}] \neq \zeta_{\cdot \beta(\omega)}$ and $\E[\zeta_{\cdot \beta(\omega)}^{(2)}|\mathcal{D}] \neq \zeta_{\cdot \beta(\omega)}$, while $\E[\zeta_{\cdot \beta(\omega)}^{(3)}|\mathcal{D}] = \zeta_{\cdot \beta(\omega)}$ as desired. Details are presented in Algorithm \ref{alg:DRSGEE2}.
	
	\subsection{Exploiting sparsity}
	S-IPW-GEE2 and S-DR-GEE2 in their current forms are not any faster than their deterministic counterparts. Rather, the convenient matrix notation in Eqs \ref{SIPWGEE2} and \ref{SDRGEE2} obscures the fact that $W^R_{i(\omega)}$ is a diagonal matrix, so one need not perform the standard matrix multiplication but rather resort to vectorized operations. The stochastic $\widetilde{W}^R_{i(\omega)}$ not only is diagonal, but also encompasses many zeros along its diagonal for which we can further exploit sparsity operations. 
	
	More formally, for a $b\times b$ diagonal matrix $\Lambda$, $a\times b$ matrix $M$, and $b\times c$ matrix $N$, computing $M(\Lambda N)$ through schoolbook matrix multiplication would have total complexity $\mathcal{O}(b^2c + abc)$. But, most of these computations are redundant, since they involve multiplying or adding zero. Denote $\Lambda'$ as the $b' \times b'$ diagonal matrix with the zero diagonal entries of $\Lambda$ removed, and denote $\lambda', \lambda$ as the vectorizations of the diagonal entries of $\Lambda', \Lambda$, respectively. Define $\text{col}_{\lambda}: \mathbb{R}^{b\times b} \rightarrow \mathbb{R}^{b\times b'}$ as the function which removes the columns of its input corresponding the zero entries of $\lambda$, and $\text{row}_{\lambda}: \mathbb{R}^{b\times b} \rightarrow \mathbb{R}^{b'\times b}$ similarly for the rows. Then we see that $M(\Lambda N) = \text{col}_\lambda(M)(\lambda'\circ \text{row}_\lambda(N))$, where $\circ$ denotes the Hadamard product, yet the complexity of $\text{col}_\lambda(M)(\lambda'\circ \text{row}_\lambda(N))$ through schoolbook matrix multiplication is $\mathcal{O}(b'c + ab'c)$. Relating back either S-IPW-GEE2 or S-DR-GEE2, the induced IPW matrices $\widetilde{W}_{\cdot i (\omega)}^{R}$ and $\widetilde{W}_{\cdot i (\omega)}^{R'}$ play the role of $\Lambda$, hence motivating our subsampling schemes where $b' \ll b$ to greatly improve iteration speed. The bottleneck in computation lies with the working correlation structure. We summarize time complexity results in the Theorem below.  \\
	\textbf{Theorem:}  \textit{Let $\pi_S \sim (\max_i n_i)^{-1}$. In the presence of standard Fisher scoring, an iteration of the GEE1 portion with 
	\begin{enumerate}
		\item[(i)] arbitrary correlation matrix
		\item[(ii)] equicorrelation matrix
		\item[(iii)] no correlation are of complexities
	\end{enumerate} are of complexities (i) $\mathcal{O}(\max_i n_i^3)$, (ii) $\mathcal{O}(\max_i n_i)$, and (iii) $\mathcal{O}(\max_i n_i)$. Similarly, standard Fisher scoring on the GEE2 portion yields (i) $\mathcal{O}(\max_i n_i^6)$, (ii) $\mathcal{O}(\max_i n_i^2)$, and (iii) $\mathcal{O}(\max_i n_i^2)$; stochastic Fisher scoring on the GEE1 portion yields (i) $\mathcal{O}(\max_i n_i^3)$, (ii) $\mathcal{O}(\max_i n_i)$, and (iii) $\mathcal{O}(1)$; stochastic Fisher scoring on the GEE2 portion yields (i) $\mathcal{O}(\max_i n_i^6)$, (ii) $\mathcal{O}(\max_i n_i^2)$, and (iii) $\mathcal{O}(1)$.}
	
	See proofs in Appendix \ref{timecomplexproofs}. Table \ref{timecomplex} expresses a clearer schematic of the Theorem, with the addition of the identity covariance structure as a special case of independence covariance structure. These time complexities are true for all of TMEE, OMEE, and PSEE; hence for the rest of this section, we refer to just full or stochastic GEE2. 
	
	\spacingset{1}
	\begin{table*}[!htbp]\centering
		\ra{1.3}
		\begin{tabular}{@{}rrrcrr@{}}\toprule
			& \multicolumn{2}{c}{Full} & \phantom{abc}& \multicolumn{2}{c}{Stochastic} \\
			\cmidrule{2-3} \cmidrule{5-6}
			& GEE1 portion & GEE2 portion  && GEE1 portion & GEE2 portion  \\ \midrule
			Arbitrary structure & $\mathcal{O}(\max_i n_i^3)$ & $\mathcal{O}(\max_i n_i^6)$  && $\mathcal{O}(\max_i n_i^3)$ & $\mathcal{O}(\max_i n_i^6)$  \\
			Equicorrelated  & $\mathcal{O}(\max_i n_i)$ & $\mathcal{O}(\max_i n_i^2)$  && $\mathcal{O}(\max_i n_i)$ & $\mathcal{O}(\max_i n_i^2)$ \\
			Independence  & $\mathcal{O}(\max_i n_i)$ & $\mathcal{O}(\max_i n_i^2)$  && $\mathcal{O}(1)$ & $\mathcal{O}(1)$  \\
			Identity  & $\mathcal{O}(\max_i n_i)$ & $\mathcal{O}(\max_i n_i^2)$  && $\mathcal{O}(1)$ & $\mathcal{O}(1)$  \\
			\bottomrule
		\end{tabular}
		\caption{Time complexities for SGEE2 algorithms under various working covariance structures.}\label{timecomplex}
	\end{table*}
	\spacingset{1.45}
	
	If we choose to model with equicorrelated $\rho_{ijj'} = \rho_{i}$, as commonly done in CRT's \citep{hayes2009cluster,crespi2009} and assume identity working correlation for the GEE2 portion in both cases, then the full GEE2 would have $\mathcal{O}(\max_i n_i)$ for the GEE1 portion and $\mathcal{O}(\max_i n_i^2)$ for the GEE2 portion, hence the overall complexity is $\mathcal{O}(\max_i n_i^2)$. With SGEE2, while the GEE1 portion remains at $\mathcal{O}(\max_i n_i)$, the GEE2 portion now becomes $\mathcal{O}(1)$, and hence SGEE2 has overall complexity of $\mathcal{O}(\max_i n_i)$. Therefore, SGEE2 cuts down the computation per iteration from roughly a quadratic rate to roughly a linear rate. If we allow the GEE1 portion to also have an independence correlation structure, then the effect of SGEE2 is even more dramatic, cutting complexity from $\mathcal{O}(\max_i n_i^2)$ to $\mathcal{O}(1)$. Additionally, SGEE2 is endowed with two more advantages. Firstly, as mentioned before, the noisier gradient calculated at each step is more likely to jerk the algorithm out of divergence due to, say, a poor initialization. Secondly, again due to sparsity, we require far less memory allocation. With full GEE2, all $\binom{n_i+1}{2}$ entries of the $E_i$ matrix would need to be stored, while SGEE2 requires $\binom{\upsilon_i+1}{2}$ entries. Since $\pi_S \sim (\max_i n_i)^{-1}$, $\upsilon_i$ is bounded, the number of entries needed to be stored does not increase with respect to $n_i$.
	\subsection{Par-SGEE2} \label{sec:parsgee2}
	While SGEE2 algorithms allow faster computations in its iterative fitting procedure, each iteration is not as informative due to the variation from the induced missingness. Hence, more iterations of SGEE2 would be needed in order to solve the estimating equations, although in practice the additional time in running more iterations is far less significant than the computational savings per iteration. Nevertheless, in pursuit of a SGEE2 variant requiring fewer iterations, we propose the Parallel SGEE2 (Par-SGEE2) class of algorithms. The general technique of parallelized SGD is expanded upon in \cite{zinkevich2010}, and one specific example applied on S-DR-GEE2 is given in Algorithm \ref{alg:ParSGEE2} in Appendix \ref{appendixalgo}. The basic idea is, after sufficiently enough iterations of SGEE2, the stochastic estimates will become unbiased and further iterations are meant to reduce variation from the stochastic nature of the algorithm. Rather, one can run $K$ independent chains of SGEE2 and average the resulting convergent estimates. Both running more iterations on a single chain or averaging over multiple chains has the same effect in reducing the variation in estimates, but with the former, the iterations must be done sequentially and hence the user must wait, while with the latter, the chains can be run in parallel. 
	
	As discussed in Section \ref{sec:sgee2}, SGEE2 reduces the frequency of divergence, but generally not all of it; there remains a non-negligible probability that the algorithm will diverge. Par-SGEE2 inherently solves the convergence issue because at least some of the chains would have converged. The average of these convergent solutions is one estimator, or better yet, one can then feed this estimator as an initial value on another run of Par-SGEE2, since the provided estimate would act as a better initial starting value and reduce the number of divergences. In a sense, Par-SGEE2 is very similar to multistart search because each chain initially fluctuates around the search space, effectively acting as a scattering of starting values. At the same time, this scattering is informative because each chain is still trying to fit on a subset of data. Hence, Par-SGEE2 offers an advantage in intrinsically incorporating information in its multistart search rather than truly random scattering.
	
	\section{Simulation} \label{sec:simulation}
	We perform two sets of experiments. The first set explores the statistical properties of IPW-GEE2 and DR-GEE2 under combinations of correctly specified / misspecified PS model and correctly specified / misspecified OM, all of which include the ICC estimates embedded in the working correlation structure in the GEE1 portion. We include analogous estimates from a parametric mixed effects model and GEE1 with independence working correlation structure for comparison, as per the discussion in Section \ref{sec:inference}. In the second set of simulations, we compare the algorithmic properties (convergence \& run-time) of stochastic DR-GEE2 and standard DR-GEE2 under various cluster size / number of cluster combinations. 
	
	We consider the following two data generation processes for binary data $Y_{ij}$ (or $R_{ij}$):
	\begin{equation} \label{truth}
	\begin{aligned} 
	\text{Parzen's method}
	&\begin{cases}
	\logit(\pi_{ij}) &= (\beta_{0Y} + \beta_{0AY} A_i) + (\bb_{ZY} + \bb_{ZAY} A_i)^\intercal \textbf{Z}_{i}  \\
	&\qquad+ (\bb_{XY} + \bb_{XAY} A_i)^\intercal \textbf{X}_{ij} \\
	\text{atanh}(\rho_{i}) &= (\alpha_{0Y} + \alpha_{0AY} A_i) + (\ba_{ZY} + \ba_{ZAY} A_i)^\intercal \textbf{Z}_{i} \\
	(\mathfrak{L}_i, \mathfrak{U}_i) &= \left(-\sqrt{\frac{\min(\bpi_i)}{1 - \min(\bpi_i)}}, \sqrt{\frac{1-\max(\bpi_i)}{\max(\bpi_i)}}\right)\\
	(\delta_i, \epsilon_i) &= \left(\frac{\mathfrak{U}_i(-\mathfrak{U}_i\mathfrak{L}_i-\rho_i)}{(\mathfrak{U}_i-\mathfrak{L}_i)\rho_i}, \frac{-\mathfrak{L}_i(-\mathfrak{U}_i\mathfrak{L}_i-\rho_i)}{(\mathfrak{U}_i-\mathfrak{L}_i)\rho_i}\right) \\
	\xi_i|A_i, \textbf{Z}_i &\sim (\mathfrak{U}_i - \mathfrak{L}_i)\text{Beta}(\delta_i, \epsilon_i) + \mathfrak{L}_i \\
	Y_{ij}|A_i, \textbf{Z}_i, \textbf{X}_i, \xi_i &\sim \text{Bernoulli}\left(\pi_{ij} + \xi_i  \sqrt{\pi_{ij}(1-\pi_{ij})}\right) \end{cases}  \\
	\text{Random intercept}
	&\begin{cases}
	\logit(\pi_{ij}) &= (\beta_{0Y} + \beta_{0AY} A_i) + (\bb_{ZY} + \bb_{ZAY} A_i)^\intercal \textbf{Z}_{i}  \\
	&\qquad+ (\bb_{XY} + \bb_{XAY} A_i)^\intercal \textbf{X}_{ij} \\
	\xi_i|A_i &\sim N(0, (\frac{1}{3} + \frac{1}{2}A_i)^2) \\
	\logit(p_{ij}) &= \xi_i + \logit(\pi_{ij}) \\
	Y_{ij}|A_i, \textbf{Z}_i, \textbf{X}_i, \xi_i &\sim \text{Bernoulli}\left(p_{ij}\right) \end{cases}
	\end{aligned}
	\end{equation}
	Parzen's method \citep{parzen2009} offers a random-effects form that attains nominal levels of $\pi_{ij}$ and $\rho_{i}$ (i.e. $\Prob(Y_{ij}|A_i, \textbf{Z}_i, \textbf{X}_i) = \pi_{ij}$ and $\corr(Y_{ij}, Y_{ij'}|A_i, \textbf{Z}_i) = \rho_{i})$ and specifically generates equicorrelated data. To ensure $0 \le  \pi_{ij} + \xi_i  \sqrt{\pi_{ij}(1-\pi_{ij})} \le 1$, one must ensure that $-\mathfrak{U}_i\mathfrak{L}_i-\rho_i \ge 0$ for all $i$. The random intercept is the traditional approach in inducing correlation among observations in a cluster. With a normal random intercept, the marginal probability of success
	\begin{equation}\label{logisticmarg}
	\begin{aligned}
	\Prob(Y_{ij} = 1|A_i, \textbf{Z}_i, \textbf{X}_i) = \int \Prob(Y_{ij} = 1|\xi_i, A_i, \textbf{Z}_i, \textbf{X}_i) dP(\xi_i) = \int \frac{e^{\xi_i + L(\bb; A_i, \textbf{Z}_i, \textbf{X}_i)}}{1+e^{\xi_i + L(\bb; A_i, \textbf{Z}_i, \textbf{X}_i)}} dP(\xi_i)
	\end{aligned}
	\end{equation}
	where $L(\bb; A_i, \textbf{Z}_i, \textbf{X}_i)$ is the linear function, is not of the logistic form and will not have a closed-form. Furthermore, the ICC is induced linearly on the logit scale, yet the manifested ICC after performing the expit function and appropriate marginalization will vary within-cluster and hence is unsuitable for simulation of equicorrelated data. We use Parzen's method to generate the ideal case of equicorrelated outcomes, while we use random intercept to induce non-equicorrelated outcomes. Furthermore, since the normal random intercept is not of the logistic form, any OM we fit with logistic regression is necessarily a misspecified model, yet we show that the marginalization interpretation $\rho_i^* = \E[\rho_{ijj'}^\dagger|A_i]$ holds.
	\subsection{Consistency and efficiency of IPW-GEE2 \& DR-GEE2 schemes} \label{sec:consis_and_effic}
	Let $\mathcal{U}(a, b)$ denote the continuous uniform distribution on $(a, b)$, and let $\mathcal{U}\{a, b\}$ denote the discrete uniform distribution on $\{a, a+1, \cdots, b-1, b\}$. To evaluate the asymptotic properties of GEE2, we set the number of clusters to an unrealistic $I = 2000$ with cluster sizes $n_i \sim \mathcal{U}\{80, 140\}$ so that we have average cluster size $\E[n_i] = 110$. The setting with large number of clusters allows us to observe asymptotic properties more quickly and to avoid computational issues that will be explored in Section \ref{algchar}. We generate $A_i \sim \text{Ber}(1/2)$ and choose $\textbf{X}_{ij} \in \mathbb{R}^3$ and $\textbf{Z}_i \in \mathbb{R}$. Details regarding generation of $\textbf{X}_{ij}$, $\textbf{Z}_i$ and choice of coefficients for $Y_{ij}$ are presented in Table \ref{generation}. We also generate $R_{ij}$ with these same covariates and coefficients for simplicity.
	
	\spacingset{1}
	\begin{table*}[!htbp]\centering
		\begin{tabular}{c|ccccc} 
			Covariate & Intercept &\multicolumn{3}{|c|}{$\textbf{X}_{ij}$} & $\textbf{Z}_{i}$ \\ 
			\hline
			Generation & -- & $\mathcal{U}(20, 60)$ & $\mathcal{U}(1, 10)$ & $\mathcal{U}(4, 25)$ & $\mathcal{U}\{80, 140\}$ \\ 
			Main-effects $\bb_{\cdot Y}$ & 0.11 & $-0.007$ & $-0.020$ & $-0.040$ & 0.009 \\ 
			Interaction $\bb_{\cdot AY}$ & 0.67 & 0.012 &  0.030 & 0.060 & $-0.018$ \\ 
			Main-effects $\ba_{\cdot Y}$ & $-0.32$ & -- & -- & -- & 0.004 \\ 
			Interaction $\ba_{\cdot Y}$ &  0.96 & -- & -- & -- & $-0.008$ 
		\end{tabular}
		\caption{Information regarding the generation process} \label{generation}
	\end{table*}
	\spacingset{1.45}
	
	The values in Table \ref{generation} are carefully chosen to guarantee $-\mathfrak{U}_i\mathfrak{L}_i-\rho_i \ge 0$ in Parzen's method. The resulting values for $\Prob(Y_{ij} = 1|A_i, \textbf{Z}_i, \textbf{X}_i)$ and $\corr(Y_{ij}, Y_{ij'}|A_i, \textbf{Z}_i, \textbf{X}_i)$, after marginalizing out $\xi_i$, are in the range [0.324, 0.733] and [0.004, 0.306], respectively. For the random-intercept method, the values of $\Prob(Y_{ij} = 1|A_i, \textbf{Z}_i, \textbf{X}_i)$ and $\corr(Y_{ij}, Y_{ij'}|A_i, \textbf{Z}_i, \textbf{X}_i)$  are in the range [0.333, 0.738] and [0.022, 0.134], respectively. The true treatment coefficients $(\bb_Y^*, \ba_Y^*)$ in the canonical TM can be calculated by numerically integrating out all other covariates except for $A_i$ in $\pi_{ij}$ and $\rho^\dagger_{ijj'}$:
	\begin{equation}\label{num_int}
	\begin{aligned}
	\expit(\beta_{0Y}^* + \beta_{AY}^*A_i) &= \int_{\mathbb{R}^4}\pi_{ij} dP(\textbf{x}_{ij})dP(\textbf{z}_i) \\
	\tanh(\alpha_{0Y}^* + \alpha_{AY}^*A_i) &= \int_{\mathbb{R}^7}\rho^\dagger_{ijj'} dP(\textbf{x}_{ij})dP(\textbf{x}_{ij'})dP(\textbf{z}_i)
	\end{aligned}
	\end{equation}
	Under Parzen's method, we obtain the values $(\bb_Y^*, \ba_Y^*) = (0.1413, 0.1808, 0.1238, 0.0755)$, and under random intercept, we obtain $(\bb_Y^*, \ba_Y^*) = (0.1378, 0.1429, 0.0307, 0.1032)$.
	
	The results in Table \ref{ParzenmethodGEE2} display biases, replicate standard errors, and average sandwich standard errors of estimated parameters from several models with $\mathcal{R} = 1000$ replicate generations of missingness and outcome, both using Parzen's method. For the mixed effects model, we fit the following on the complete case data:
	\begin{equation}\label{mixed}
	\begin{aligned}
	\logit \{\Prob(Y_{ij}=1|A_i, \xi_i)\} &= \widetilde{\beta}_0 + \widetilde{\beta}_A A_i + \xi_i \\
	\xi_i|A_i &\sim N(0, \widetilde{\sigma}^2_{A_i})
	\end{aligned}
	\end{equation}
	which takes nearly the functional form of the random intercept generation process in Eq \ref{truth}, less the baseline covariates. Using the marginalizations in Eqs \ref{logisticmarg} and \ref{num_int}, we can obtain $(\beta_{0Y}^*, \beta_{AY}^*, \alpha_{0Y}^*, \alpha_{AY}^*)$ from $(\widetilde{\beta}_0, \widetilde{\beta}_A, \widetilde{\sigma}^2_0, \widetilde{\sigma}^2_1)$ and standard errors for $\beta_{0Y}^*, \beta_{AY}^*$ from the standard errors of $\widetilde{\beta}_0, \widetilde{\beta}_A$ through the delta method. Unfortunately, analytical standard errors for $\alpha_{0Y}^*, \alpha_{AY}^*$ require standard errors of $\widetilde{\sigma}^2_0, \widetilde{\sigma}^2_1$, for which methods are less well-developed \citep{bates2010,mcculloch2001,wu2012comparison}. Hence, while we report replicate standard errors for $\widetilde{\sigma}^2_0, \widetilde{\sigma}^2_1$, we omit sandwich error standard errors. Mixed effects models naturally handle MAR if the true generation process follows the form in Eq \ref{mixed}. Certainly, both generation processes in Eq \ref{truth} do not; Parzen's method does not follow the mixed effects framework and our random intercept method, while is a mixed effects model, incorporates additional covariates for which Eq \ref{mixed} does not.
	
	\spacingset{1}
	\begin{table*}[ht!]\centering
		\ra{1.1}
		\footnotesize{
			\begin{tabular}{@{}cccc@{}}\toprule
				& \specialcell{Averaged bias \\ (Replicate SE) \\ (Averaged sandwich SE)} && \specialcell{Averaged bias \\ (Replicate SE) \\ (Averaged sandwich SE)} \\
				\cmidrule{2-2} \cmidrule{4-4}
				& $\begin{array}{cccc}\beta_{0Y}^* \qquad \qquad & \beta_{AY}^* \qquad \qquad & \alpha_{0Y}^* \qquad \qquad & \alpha_{AY}^* \qquad \end{array}$ && $\begin{array}{cccc}\beta_{0Y}^* \qquad \qquad & \beta_{AY}^*  \end{array}$  \\ 
				\textbf{Complete Case Mixed Effects} \\ \midrule \midrule
				& $\begin{array}{cccc} 0.0421 & -0.0238 & 0.0016 & -0.0009 \\ (0.0227) & (0.0364) & (0.0053) & (0.0088) \\  (0.0238) & (0.0373) & \text{---} & \text{---} \end{array}$  && \Huge ---  \\ \\
				\textbf{GEE} & \textit{GEE2}  && \textit{GEE1} \\ 
				\midrule \midrule \\
				\textbf{Complete Case} \\ \midrule 
				& $\begin{array}{cccc} 0.0349 & -0.0239 & 0.0113 & -0.0016 \\ (0.0245) & (0.0379) & (0.0070) & (0.0121) \\  (0.0238) & (0.0380) & (0.0069) & (0.0117) \end{array}$  && $\begin{array}{cccc} 0.0413 & -0.0228
				\\ (0.0262) & (0.0404) \\  (0.0260) & (0.0416) \end{array}$  \\ 
				\textbf{PSM Correctly Specified} \\ \midrule
				$\mathcal{G}_1(\textbf{R})$ IPW & $\begin{array}{cccc} -0.0006 & 0.0020 & 0.0024 & -0.0008
				\\ (0.0257) & (0.0398) & (0.0064) & (0.0112)
				\\  (0.0249) & (0.0400) & (0.0064) & (0.0111)
				\end{array}$ && $\begin{array}{cccc} -0.0003	& 0.0010
				\\ (0.0252) & (0.0391) \\  (0.0252)& (0.0405)    \end{array}$  \\
				\cdashlinelr{2-4} 
				$\mathcal{G}_2(\textbf{R})$ IPW & $\begin{array}{cccc} -0.0005 & 0.0019 & -0.0001 & 0.0002
				\\ (0.0258) & (0.0399) & (0.0066) & (0.0112)
				\\  (0.0249) & (0.0401) & (0.0063) & (0.0109) \end{array}$  && \Huge --- \\
				\cdashlinelr{2-4} 
				Doubly-Robust & $\begin{array}{cccc} -0.0006 & 0.0018 & -0.0003 & 0.0003
				\\ (0.0262) & (0.0399) & (0.0061) & (0.0111)
				\\  (0.0297) & (0.0389) & (0.0060) & (0.0108)
				\end{array}$  && $\begin{array}{cccc} -0.0004 & 0.0010
				\\ (0.0251) & (0.0391)  \\  (0.0246) & (0.0404) \end{array}$ \\ 
				\textbf{PSM Misspecified} \\ \midrule
				$\mathcal{G}_1(\textbf{R})$ IPW & $\begin{array}{cccc} 0.0341 & -.0124 & 0.0112 & -0.0018 \\ (0.0255) & (0.0414) & (0.0068) & (0.0116) \\  (0.0255) & (0.0411) & (0.0068) & (0.0117)\end{array}$ && $\begin{array}{cccc} 0.0341 & -0.0121	\\ (0.0264) & (0.0401) \\  (0.0260) & (0.0416) \end{array}$\\
				\cdashlinelr{2-4}
				$\mathcal{G}_2(\textbf{R})$ IPW & $\begin{array}{cccc} 0.0326 & -0.0092 & 0.0089 & 0.0022
				\\ (0.0252) & (0.0411) & (0.0067) & (0.0117)
				\\  (0.0255) & (0.0411)  & (0.0067) & (0.0117) \end{array}$  && \Huge --- \\
				\cdashlinelr{2-4}
				Doubly-Robust & $\begin{array}{cccc} 0.0000 & 0.0005 & -0.0002 & -0.0001
				\\ (0.0251) & (0.0401)	& (0.0061) & (0.0107)
				\\  (0.0303) & (0.0397) & (0.0064) & (0.0114)
				\end{array}$   && $\begin{array}{cccc} -0.0002 & 0.0007
				\\ (0.0252) & (0.0392) \\  (0.0253) & (0.0415) \end{array}$ \\
				\bottomrule
			\end{tabular}
		}
		\caption{Biases \& Standard Errors from 1000 replicate simulations with both $Y_{ij}, R_{ij}$ simulated with Parzen's method.} \label{ParzenmethodGEE2}
	\end{table*}
	\spacingset{1.45}
	
	For the IPW-GEE2 fits, we distinguish $\mathcal{G}_1(\textbf{R})$ IPW and $\mathcal{G}_2(\textbf{R})$ IPW as the IPW models with and without accounting for the correlation among the missingness indicators, respectively, as discussed in Section \ref{sec:ipwgee2}. For GEE1, there naturally is no model for correlated missingness, and that block is omitted. The fitted OM and correctly-specified PSM are 
	\begin{equation}\label{PSMcorr}
	\begin{aligned}
	\logit(\pi_{ij}) &= (\beta_{0Y} + \beta_{0AY} A_i) + (\bb_{ZY} + \bb_{ZAY} A_i)^\intercal \textbf{Z}_{i} + (\bb_{XY} + \bb_{XAY} A_i)^\intercal \textbf{X}_{ij} \\
	\text{atanh}(\rho_{ijj'}) &= (\alpha_{0Y} + \alpha_{0AY} A_i) + (\ba_{ZY} + \ba_{ZAY} A_i)^\intercal \textbf{Z}_{i}
	\end{aligned}
	\end{equation}
	i.e. the exact model used to generate $R_{ij}, Y_{ij}$ from Parzen's method. The fitted misspecified PSM is
	\begin{equation}\label{PSMincorr}
	\begin{aligned}
	\logit(\pi_{ij}) &= \beta_{0Y} + \beta_{AY} A_i + \bb_{ZY}^\intercal \textbf{Z}_{i} + \bb_{XY}^\intercal \textbf{X}_{ij} \\
	\text{atanh}(\rho_{ijj'}) &= \alpha_{0Y} + \alpha_{AY} A_i + \ba_{ZY}^\intercal \textbf{Z}_{i}
	\end{aligned}
	\end{equation}
	i.e. the model with interaction terms of $A_i$ with $\textbf{Z}_i, \textbf{X}_i$ are omitted.
	
	\spacingset{1}
	\begin{table*}[ht!]\centering
		\ra{1.1}
		\footnotesize{
		\begin{tabular}{@{}cccc@{}}\toprule
			& \specialcell{Averaged bias \\ (Replicate SE) \\ (Averaged sandwich SE)} && \specialcell{Averaged bias \\ (Replicate SE) \\ (Averaged sandwich SE)} \\
			\cmidrule{2-2} \cmidrule{4-4}
			& $\begin{array}{cccc}\beta_{0Y}^* \qquad \qquad & \beta_{AY}^* \qquad \qquad & \alpha_{0Y}^* \qquad \qquad & \alpha_{AY}^* \qquad \end{array}$ && $\begin{array}{cccc}\beta_{0Y}^* \qquad \qquad & \beta_{AY}^*  \end{array}$  \\ 
			\textbf{Complete Case Mixed Effects} \\ \midrule \midrule
			& $\begin{array}{cccc} 0.0343 & -0.0244 & -0.0005 & -0.0001 \\ (0.0144) & (0.0290) & (0.0020) & (0.0058) \\  (0.0139) & (0.0279) & \text{---} & \text{---} \end{array}$  && \Huge ---  \\ \\
			\textbf{GEE} & \textit{GEE2}  && \textit{GEE1} \\ 
			\midrule \midrule \\
			\textbf{Complete Case} \\ \midrule
			& $\begin{array}{cccc} 0.0340 & -0.0266 & -0.0005 & -0.0004 \\ (0.0143) & (0.0291) & (0.0022) & (0.0071) \\  (0.0140) & (0.0284) & (0.0022) & (0.0070)
			\end{array}$  && $\begin{array}{cccc} 0.0400	& -0.0239
			\\ (0.0145) & (0.0303) \\  (0.0143)	& (0.0299) \end{array}$  \\ 
			\textbf{PSM Correctly Specified} \\ \midrule
			$\mathcal{G}_1(\textbf{R})$ IPW & $\begin{array}{cccc} -0.0001 & -0.0020 & -0.0002 & 0.0003
			\\ (0.0148)	& (0.0295) & (0.0023) & (0.0070)
			\\  (0.0143)& (0.0297) & (0.0022) & (0.0071)
			\end{array}$ && $\begin{array}{cccc} -0.0002 & 0.0003
			\\ (0.0143) & (0.0297) \\  (0.0143)	& (0.0299) \end{array}$  \\
			\cdashlinelr{2-4} 
			$\mathcal{G}_2(\textbf{R})$ IPW & $\begin{array}{cccc} -0.0001& -0.0021 & -0.0001 & 0.0002 
			\\ (0.0150) & (0.0296) & (0.0023) & (0.0070)
			\\  (0.0143) & (0.0297) & (0.0022) & (0.0071)\end{array}$  && \Huge --- \\
			\cdashlinelr{2-4} 
			Doubly-Robust & $\begin{array}{cccc} -0.0001 & -0.0020 & -0.0001 & 0.0003
			\\ (0.0149)	& (0.0294) & (0.0023) & (0.0070)
			\\  (0.0212) & (0.0248)	& (0.0022)	& (0.0071)
			\end{array}$  && $\begin{array}{cccc} 0.0000 & 0.0003
			\\ (0.0139) & (0.0297) \\  (0.0137) & (0.0299) \end{array}$ \\ 
			\textbf{PSM Misspecified} \\ \midrule
			$\mathcal{G}_1(\textbf{R})$ IPW & $\begin{array}{cccc} 0.0328 & -0.0157 & -0.0005 & -0.0003 \\ (0.0145) & 0.0303 &(	0.0022) & (0.0071) \\  (0.0143) & (0.0297) & (0.0022) & 0.0070 \end{array}$ && $\begin{array}{cccc} 0.0327 & -0.0134
			\\ (0.0145) & (0.0302) \\  (0.0143) & (0.0299) \end{array}$\\
			\cdashlinelr{2-4}
			$\mathcal{G}_2(\textbf{R})$ IPW & $\begin{array}{cccc} 0.0313 & -0.0128 & -0.0005 & -0.0005
			\\ (0.0145) & (0.0304) & (0.0022) & (0.0071)
			\\  (0.0142) & (0.0297) & (0.0022) & (0.0071) \end{array}$  && \Huge --- \\
			\cdashlinelr{2-4}
			Doubly-Robust & $\begin{array}{cccc} -0.0006 & -0.0006 & -0.0001 & -0.0001
			\\ (0.0145) & (0.0296) & (0.0022) & (0.0070)
			\\  (0.0211) & (0.0247)	 & (0.0022) & (0.0069)
			\end{array}$   && $\begin{array}{cccc} -0.0008 & 0.0013
			\\ (0.0141)	& (0.0302) \\  (0.0137) & (0.0299)
			\end{array}$ \\
			\bottomrule
		\end{tabular}
	}
		\caption{Biases \& Standard Errors from 1000 replicate simulations with $R_{ij}$ simulated using Parzen's method and $Y_{ij}$ simulated using random-intercept method.} \label{randominterceptGEE2}
	\end{table*}
	\spacingset{1.45}
	
	The following discussion in comparing the performance of each estimation procedure is based on the replicate Wald statistic $W = \sqrt{\mathcal{R}}\cdot \frac{\text{Bias}}{\text{Std Error}}$ and checking whether $|W| > 2$. Using this metric and the information from Table \ref{ParzenmethodGEE2}, when PSM is correctly specified, complete case analysis (for both mixed effects, GEE1, and GEE2) leads to severe bias in estimating all parameters. $\mathcal{G}_1(\textbf{R})$ IPW-GEE2  and IPW-GEE1 provide consistent estimates for the mean parameters $\beta_{0Y}^*$ and $\beta_{AY}^*$, although the former still fails to correctly estimate the association parameters $\alpha_{0Y}^*$ and $\alpha_{AY}^*$.  $\mathcal{G}_2(\textbf{R})$ IPW-GEE2 and doubly-robust GEE2 and GEE1 produce consistent estimates for all parameters estimable under their respective models. When PSM is misspecified, we note that only DR-GEE2 and DR-GEE1 produce consistent estimates. Note that the sandwich variance estimators in general are close to the true sampling variance with the exception of $\beta_{0Y}$ under the DR-GEE2 model, for which it is somewhat conservative. We also observe that DR-GEE1 (with independence correlation structure) standard errors of the mean parameters $\beta_{0Y}^*, \beta_{AY}^*$ are smaller than the DR-GEE2 standard errors of $\beta_{0Y}^*, \beta_{AY}^*$. 
	
	The results in Table \ref{randominterceptGEE2} display biases, replicate standard errors, and sandwich standard errors of estimated parameters from several models with $\mathcal{R} = 1000$ replicate generations of missingness using Parzen's method and outcome using random intercepts. We still fit the correct OM and PSM  using Eq \ref{PSMcorr} and incorrect PSM using Eq \ref{PSMincorr}. Note that the true OM is no longer of the logistic form, and hence the fitted OM will be misspecified. Nevertheless, we reach nearly identical conclusions regarding the validity of models as done with Table \ref{ParzenmethodGEE2}. Especially noteworthy is that, even when the PSM is misspecified, the DR-GEE2 produces consistent estimates of all its parameters. Consistent estimation of the mean parameters may be due to the fact that random intercept generation is still ``linear enough" with respect to the covariates. Consistent estimation of the association parameters is a bit more surprising, because it ultimately means that, even when the outcome ICC is non-equicorrelated, we may still model it with an equicorrelated OM and still produce roughly consistent estimates of the treatment ICC.
	
	\subsection{Algorithmic Characteristic of DR-GEE2 vs S-DR-GEE2} \label{algchar}
	Having established the consistency of DR-GEE2, in our second set of experiments we now compare against S-DR-GEE2. We generate both missingness and outcome using Parzen's method and the information from Table \ref{generation}, and we fit with both PSM and OM correctly specified. We now vary the number of cluster $I$ and cluster sizes $n_i$, and consider the following three scenarios: $(I, \E[n_i]) = (30, 30), (300, 30), (30, 300)$. Because the termination condition for stochastic methods based on error thresholds are a bit uncertain, since it's possible to choose a subset that, by chance, gives a very low error, we decide \textit{a prior} on the number of iterations. For S-DR-GEE2, under the scenarios with expected cluster size 30, we run $\omega = 20$ iterations to fit the PSM and OM and $\omega = 10$ iterations to fit the treatment model with sampling proportion $\pi_S = 0.30$. For the scenario with expected cluster size 300, we run $\omega = 25$ iterations to fit the PSM and OM and $\omega = 12$ iterations to fit the treatment mode with sampling proportion $\pi_S = 0.15$ and learning rates $\gamma_\omega = (\omega+1)^{-1}$. Tables \ref{full_vs_stoch} and \ref{comp_and_conv} present the statistical and algorithmic results, respectively, of DR-GEE2 and S-DR-GEE2.
	
		\spacingset{1}
		\begin{table*}[!ht]\centering
			\ra{1.1}\small{
			\begin{tabular}{@{}rccc@{}}\toprule
				\textbf{Scenarios}& \specialcell{\textbf{Full DR-GEE2} \\ Averaged bias \\ (Replicate SE) \\ (Averaged sandwich SE)} && \specialcell{\textbf{S-DR-GEE2} \\ Averaged bias \\ (Replicate SE) \\ (Averaged sandwich SE)} \\
				\cmidrule{2-2} \cmidrule{4-4}
				& $\begin{array}{cccc}\beta_{0Y}^* \qquad \qquad & \beta_{AY}^* \qquad \qquad & \alpha_{0Y}^* \qquad \qquad & \alpha_{AY}^* \qquad \end{array}$ && $\begin{array}{cccc}\beta_{0Y}^* \qquad \qquad & \beta_{AY}^* \qquad \qquad & \alpha_{0Y}^* \qquad \qquad & \alpha_{AY}^* \qquad \end{array}$  \\ \midrule
				$(I, \E[n_i]) = (30, 30)$ & $\begin{array}{cccc} 0.0067 & -0.0082 & -0.0153 & 0.0010 
				\\ (0.2563) & (0.3973) & (0.0629) & (0.1140) \\ ( 0.2541) & (0.3516) & (0.0535) & (0.0983)
				\end{array}$  && $\begin{array}{cccc} 0.0025 & 0.0071 & -0.0041 & -0.0095 
				\\ (0.2724) & (0.4084) & (0.0715) & (0.1203)
				\\ (0.2533) & (0.3513) & (0.0580) & (0.1012)
				\end{array}$  \\
				\cdashlinelr{2-4} 
				$(I, \E[n_i]) = (300, 30)$ & $\begin{array}{cccc} -0.0004 & -0.0004 & -0.0021 & 0.0004 
				\\ (0.0707) & (0.1144) & (0.0199) & (0.0338) \\ ( 0.0840) & (0.1106) & (0.0199) & (0.0339)
				\end{array}$ && $\begin{array}{cccc} 0.0015 & 0.0046 & -0.0009 & -0.0002 
				\\ (0.0759) & (0.1188) & (0.0218) & (0.0362) \\ ( 0.0842) & (0.1109) & (0.0201) & (0.0339)
				\end{array}$ \\
				\cdashlinelr{2-4} 
				$(I, \E[n_i]) = (30, 300)$ & $\begin{array}{cccc} -0.0005 & 0.0034 & -0.0124 & -0.0010 
				\\ (0.2103) & (0.3364) & (0.0552) & (0.1033)
				\\  (0.2155) & (0.2970) & (0.0388) & (0.0782)\end{array}$  && $\begin{array}{cccc} -0.0051 & 0.0067 & -0.0083 & -0.0029 
				\\ (0.2141) & (0.3486) & (0.0468) & (0.0872)
				\\  (0.2170) & (0.2952) & (0.0388) & (0.0737)\end{array}$ \\
				\bottomrule
			\end{tabular}}
			\caption{Comparison of statistical and computational characteristics of full DR-GEE2 vs S-GEE2. $\mathcal{R} = 2000$ replicate simulations.}\label{full_vs_stoch} 
		\end{table*}
		\begin{table*}[!ht]\centering
			\ra{1.3}\scriptsize{
			\begin{tabular}{@{}rccccc@{}}\toprule
				& \texttt{geese} && \textbf{Full DR-GEE2} && \textbf{S-DR-GEE2} \\
				\cmidrule{2-2} \cmidrule{4-4} \cmidrule{6-6} 
				\textbf{$(I, \E[n_i])$} & $\begin{array}{cccc} (30, 30) & (300, 30) & (30, 300) \end{array}$ && $\begin{array}{cccc} (30, 30) & (300, 30) & (30, 300) \end{array}$ && $\begin{array}{cccc} (30, 30) & (300, 30) & (30, 300) \end{array}$ \\
				\textbf{Convergence} \\ \midrule
				\% PSM error only & $\begin{array}{cccc} \text{---} \quad \quad \quad & \text{---} \quad \quad \quad & \text{---} \end{array}$ && $\begin{array}{cccc} 4.22\% \quad \quad & 0.41\% \quad \quad & 7.97\% \end{array}$ && $\begin{array}{cccc} 0.58\% \quad \quad & 0.10\% \quad \quad & 1.68\% \end{array}$ \\ 
				\% OM error only & $\begin{array}{cccc} \text{---} \quad \quad \quad & \text{---} \quad \quad \quad & \text{---} \end{array}$ && $\begin{array}{cccc} 9.03\% \quad \quad & 0.86\% \quad \quad & 11.80\% \end{array}$ && $\begin{array}{cccc} 9.38\% \quad \quad & 0.77\% \quad \quad & 6.30\% \end{array}$ \\
				\% PSM or OM error & $\begin{array}{cccc} \text{---} \quad \quad \quad & \text{---} \quad \quad \quad & \text{---} \end{array}$ && $\begin{array}{cccc} 0.36\% \quad \quad & 0.00\% \quad \quad & 0.49\% \end{array}$ && $\begin{array}{cccc} 0.12\% \quad \quad & 0.00\%\quad \quad & 0.11\% \end{array}$\\
				\% Conditional TM error & $\begin{array}{cccc} 0\% \quad \quad \quad & 0\% \quad \quad \quad & 26\% \end{array}$ && $\begin{array}{cccc} 2.13\% \quad \quad & 0.00\% \quad \quad & 3.97\% \end{array}$ && $\begin{array}{cccc} 1.23\% \quad \quad & 0.00\% \quad \quad & 0.41\% \end{array}$ \\ \\
				\textbf{\specialcell{Run-time (sec)$^\dagger$}} \\ \midrule
				PSM fitting & $\begin{array}{cccc} \text{---} \quad \quad \quad & \text{---} \quad \quad \quad & \text{---} \end{array}$ && $\begin{array}{cccc} 0.38 \quad \quad & 3.88 \quad \quad & 25.69 \end{array}$ && $\begin{array}{cccc} 0.29 \quad \quad & 2.84 \quad \quad & 1.76 \end{array}$ \\ 
				OM fitting & $\begin{array}{cccc} \text{---} \quad \quad \quad & \text{---} \quad \quad \quad & \text{---} \end{array}$ && $\begin{array}{cccc} 0.20 \quad \quad & 2.05 \quad \quad & \text{ } 8.01 \end{array}$ && $\begin{array}{cccc} 0.25 \quad \quad & 2.33 \quad \quad & 0.81 \end{array}$ \\ 
				TM fitting & $\begin{array}{cccc} \text{0.10} \quad \quad \quad & \text{0.86} \quad \quad & \text{1174} \end{array}$ && $\begin{array}{cccc} 0.40 \quad \quad & 4.24 \quad \quad & 27.59 \end{array}$ && $\begin{array}{cccc} 0.31 \quad \quad & 3.14 \quad \quad & 1.53 \end{array}$ \\ 
				\hline
			\end{tabular}}
			\caption{Algorithmic analysis of standard and stochastic DR-GEE2. $\mathcal{R} = 2000$ replicate simulations. Run-time values are computed on runs which converged. The conditional TM error is the error rate among simulations whence PSM and OM converged. \\ $^\dagger$ Each replicate simulation was executed in R on a dual-core node on the Orchestra cluster supported by the Harvard Medical School Research Information Technology Group.}\label{comp_and_conv}
		\end{table*}
		\spacingset{1.45}
	
From Table \ref{full_vs_stoch}, and using the Wald statistic metric to evaluate model validity, the association parameters from the $I = 30$ sub-experiments all are biased. This is readily explained by the fact that the asymptotics for the association parameters depend on $I$ rather than $\sum_{i=1}^{I}n_i$, and hence at these small number of clusters, asymptotics haven't fully kicked in. Other than this, overall, the parameter estimates and standard errors are very similar between DR-GEE2 and S-DR-GEE2, albeit the standard errors under S-DR-GEE2 are slightly higher. This slightly higher variability can be done away with by simply asking for a few more iterations. Even so, at a small cost of higher variability, the computational savings of S-DR-GEE2 are apparent. From Table \ref{comp_and_conv}, even at small cluster sizes, which S-DR-GEE2 was not designed to be optimal, we still see moderately higher convergent solutions and somewhat less time to fit each model. We see these results further accentuated when expected cluster size is 300. Here, for all of OM, PSM, and TM, we see that S-DR-GEE2 provides up to 80\% reduction in returned errors (i.e. divergence, large condition numbers of Hessians) and approximately 90\% reduction in run-time. 
	 
We also fit a complete-case TM in each replicate simulation using the \texttt{geese} command from the \texttt{geepack} package. We see that \texttt{geese} fits faster than our algorithms in the (30, 30) and (300, 30) cases, while our code runs far faster and leads to fewer errors in the (30, 300) case. Granted, the comparisons are not the most commensurate: \texttt{geese} performs all calculations in the C programming and wraps the results into R, while our implementation is fully in R, not to mention the additional time in incorporating the IPW or DR portions. On the other hand, our use of \texttt{geese} specifies a custom correlation structure for each cluster to handle the different treatment arms, while our implementation fully exploits analytical inverses of the equicorrelation structure. 

	\section{Application to Sanitation Data} \label{sec:sanitation}
	\cite{guiteras2015} investigated the efficacy of alternative policies in encouraging use of hygienic latrines in developing countries. A total of 380 communities in rural Bangladesh were assigned to different marketing interventions -- community motivation, subsidies, supply-side market, a combination of the three and a control group. Results based on a mixed-effect model suggested supply-side market alone did not increase hygienic latrine ownership (+0.3\% points, $p$-value = 0.90). We reanalyzed this dataset with GEE2 approaches assuming that the outcome are rMAR, letting $A_i = 1$ for supply-side market alone and $A_i = 0$ for control group. We excluded all observations with missing covariates, given the low rate at which they were missing ($<1\%$). The final dataset contains 4768 individuals across 100 clusters with ten individual-level covariates (report diarrhea indicator $X_1$, male indicator $X_2$, age $X_3$, education indicator $X_4$, Muslim indicator $X_5$, Bengali indicator $X_6$, agricultor indicator $X_7$, stove indicator $X_8$, water pipes indicator $X_9$, phone indicator $X_{10}$) and five (excluding marketing intervention) cluster-level covariates (village population $Z_1$, \# of doctors $Z_2$, \% landless $Z_3$, \% almost landless $Z_4$, \% access electricity $Z_5$).
	
	\spacingset{1}
	\begin{table*}[!htb]\centering
		\ra{1.3}
		\footnotesize{
			\begin{tabular}{@{}rccccccc@{}}\toprule
				& Estimates && Sandwich SE && $p$-value && Run-time (sec)$^\dagger$ \\
				\cmidrule{2-2} \cmidrule{4-4} \cmidrule{6-6} \cmidrule{8-8}
				& $\begin{matrix} \beta_{AY}^* & \alpha_{0Y}^* & \alpha_{AY}^* \end{matrix}$ && $\begin{matrix} \beta_{AY}^* & \alpha_{0Y}^* & \alpha_{AY}^* \end{matrix}$ && $\begin{matrix} \beta_{AY}^* & \alpha_{0Y}^* & \alpha_{AY}^* \end{matrix}$ && $\begin{matrix} \text{PS} & \text{OM} & \text{TM} \end{matrix}$\\ \midrule
				CC GEE2 & $\begin{matrix} 0.207 & 0.090 & 0.015 \end{matrix}$  
				&& $\begin{matrix} 0.151 & 0.016 & 0.029 \end{matrix}$ && $\begin{matrix} 0.17 & < 0.01 & 0.60 \end{matrix}$ && $\begin{matrix} \text{ --- }\text{ } & \text{ --- }\text{ } & 1.06 \end{matrix}$ \\
				\hline 
				$\mathcal{G}_1(\textbf{R})$ IPW-GEE2 & $\begin{matrix} 0.198 & 0.090 & 0.014 \end{matrix}$  
				&& $\begin{matrix} 0.151 & 0.016 & 0.029 \end{matrix}$  && $\begin{matrix} 0.19 & < 0.01 & 0.62 \end{matrix}$ && $\begin{matrix} \text{0.10}\text{ } & \text{ --- }\text{ } & 4.39 \end{matrix}$ \\
				\hline
				$\mathcal{G}_2(\textbf{R})$ IPW-GEE2 & $\begin{matrix} 0.204 & 0.089 &  0.015 \end{matrix}$  
				&& $\begin{matrix} 0.151 & 0.016 & 0.029 \end{matrix}$  && $\begin{matrix} 0.18 & < 0.01 & 0.60 \end{matrix}$ && $\begin{matrix} 3.19^* & \text{ --- }\text{ } & 4.02 \end{matrix}$\\
				\hline
				DR-GEE2 & $\begin{matrix} 0.457 & 0.098 &  0.003 \end{matrix}$  
				&& $\begin{matrix} 0.093 & 0.016 & 0.022 \end{matrix}$ && $\begin{matrix} <0.01 & < 0.01 & 0.89 \end{matrix}$ && $\begin{matrix} 3.19^* & 3.09^* & 5.49 \end{matrix}$\\
				\hline
				\multicolumn{8}{l}{\textbf{TM:} $\logit(\pi_{i}^*) =  \beta_{0Y}^* + \beta_{AY}^*A_i$} \\
				\multicolumn{8}{l}{\textcolor{white}{\textbf{TM:}} $\text{atanh}(\rho_i^*) =  \alpha_{0Y}^* + \alpha_{AY}^*A_i$} \\
				\multicolumn{8}{l}{\textbf{PSM:} $\logit(\pi_{ij}^R) =  \beta_{0R} + \beta_{AR}A_i + \sum_{k \in \{2, 3, 5, 6, 7, 8, 10\}}\beta_{X R}^{(k)}X_{ijk} + \sum_{k \in \{1,2,3,4\}}\beta_{Z R}^{(k)}Z_{ik}$} \\
				\multicolumn{8}{l}{\textcolor{white}{\textbf{PSM:}} \qquad \qquad \qquad \quad $+ A_i\sum_{k \in \{5,6,8\}}\beta_{AX R}^{(k)}X_k + A_i\sum_{k \in \{2,3,4\}}\beta_{AZ R}^{(k)}Z_{ik}$} \\
				\multicolumn{8}{l}{\textcolor{white}{\textbf{PSM:}} $\text{atanh}(\rho_i^R) =  \alpha_{0R} + \alpha_{AR}A_i+ \sum_{k \in \{1,2,3,4\}}\alpha_{Z R}^{(k)}Z_{ik} + A_i\sum_{k \in \{2,3,4\}}\alpha_{AZ R}^{(k)}Z_{ik} $} \\
				\multicolumn{8}{l}{\textbf{OM:} $\logit(\pi_{ij}) =  \beta_{0Y} + \beta_{AY}A_i + \sum_{k \in \{1,2,3,4,5,8,9,10\}}\beta_{X Y}^{(k)}X_{ijk} + \sum_{k \in \{1,2,3,4,5\}}\beta_{Z Y}^{(k)}Z_{ik}$} \\
				\multicolumn{8}{l}{\textcolor{white}{\textbf{OM:}} \qquad \quad \qquad \qquad $+ A_i\sum_{k \in \{1,3,8\}}\beta_{AX Y}^{(k)}X_k + A_i \beta_{AZ Y}^{(5)}Z_{i5}$} \\
				\multicolumn{8}{l}{\textcolor{white}{\textbf{OM:}} $\text{atanh}(\rho_i) =  \alpha_{0Y} + \alpha_{AY}A_i+ \sum_{k \in \{1,2,3,4,5\}}\alpha_{Z Y}^{(k)}Z_{ik} + A_i \alpha_{AZ Y}^{(5)}Z_{i5} $} 
			\end{tabular}}
			\caption{Effects of the supply side-market vs. control on the probability of hygienic latrine
				ownership in the sanitation data analysis \citep{guiteras2015} using the complete-case GEE2,
				IPW-GEE2 adjustment (non-adjusting and adjusting for missingness ICC), and DR-GEE2, assuming outcomes are rMAR.\\
				$^*$ Fitted with 50 parallel stochastic GEE2, and averaging convergent estimates. Reported are median times among convergent estimates.\\
				$^\dagger$ Executed in R on a desktop with Intel(R) Core(TM) i5-4460 CPU \@3.20GHz} \label{Sanitation}
		\end{table*}
		\spacingset{1.45}
		
	Table \ref{Sanitation} present results upon fitting complete-case, $\mathcal{G}_1$ IPW, $\mathcal{G}_2$ IPW, and DR GEE2. Variables selected for the PSM and OM of the main effects were determined by backward stepwise logistic regression based on AIC, where the full model is a linear function of all covariates and the interactions terms between market intervention and all other covariates. We include all selected cluster-level covariates in the PSM and OM for the ICC (see Table \ref{Sanitation}).  We experienced convergence issues in fitting the PSM and OM to the data when using full GEE2. To overcome this, we fitted 50 parallel stochastic GEE2 (described in Section \ref{sec:parsgee2}), and averaged the convergent estimates. Complete-case and IPW-GEE2 analysis suggest similar non-significant supply-side effect (log OR $\approx$ 0.20, $p$-value $\approx$ 0.18 in all cases), but DR-GEE2 provides evidence of a significant effect (log OR = 0.46, $p$-value $< 0.01$). The propensity scores among non-missing control-group subjects are within the range [0.745, 1.000] with mean 0.964 and among the non-missing supply-side intervention group subjects are within the range [0.621, 0.995] with mean 0.956. Due to the approximate constancy and balance of the PS within both groups, the IPW-adjustment offers minor reweighing of observations and no tangible change in estimates.
This could be due to small proportion of missingness (about 3.5\%), data are missing completely at random, or the PS model is misspecified (missing important covariates or the functional form of the covariates may be misspecified). DR-GEE2 provides protection against misspecification of the PS model through augmentation. We would expect that DR-GEE2 provide consistent estimates if the OM is correctly specified. The OM suggests that households with higher education and economic status (through more stoves, water pipes, and phones) are more likely to have a hygienic latrine. Incorporating covariates that are associated with the outcome is expected to improve the efficiency of the estimation of intervention effects.  All methods conclude that there is significant treatment-specific ICC within clusters e.g. ICC$_{\text{Control}} = \tanh(0.098) \approx 0.098$ and ICC$_{\text{Supply Side}} = \tanh(0.101) \approx 0.101$ from the DR-GEE2, each with $p$-value $< 0.01$.  As none of the methods finds evidence of different treatment-specific ICC's between supply-side and control group ($p$-values = 0.60, 0.62, 0.60, 0.89), we also estimate an overall ICC of about 10\%.

	\section{Discussion} \label{sec:discusion}

In this paper, we proposed DR-GEE2 for estimating the marginal treatment effect and treatment-specific ICCs in cluster randomized trials. Our estimators are most useful in the settings where estimation of ICCs is the focus. If the interest is solely on the treatment effect on the outcomes, using working independence correlation matrix is an attractive approach due to its high efficiency in many settings and its simplicity in avoiding the need to estimate high-order association parameters. In the absence of missing data, standard GEE2 is highly efficient with a correctly specified working covariance structure. More concretely, the class of estimating functions which satisfy the canonical TM in Eq \ref{marginal} and are regular asymptotically linear (RAL) must be of the form
	\begin{align*}
	\textbf{0} &= \sum_{i=1}^{I} h(A_i)E_{i}
	\end{align*}
	The choice of index function $h(A_i) = D_i^\intercal V_i^{-1}$, which reduces back to GEE2, results in the efficient score for the canonical TM, hence attaining the minimum asymptotic variance RAL estimator for $(\bb_Y^*, \ba_Y^*)$ \citep{chamberlain1986asymptotic}. However, in the case of IPW-GEE2 or DR-GEE2, this choice is no longer optimal and the actual $h_\text{opt}(A_i)$ to achieve the efficient score is far more complicated \citep{stephens2014}. \cite{stephens2014} showed in simulation studies the efficiency gains from using $h_\text{opt}(A_i)$ are modest and very sensitive to the correct specification of all components that comprise $h_\text{opt}(A_i)$, which in practice is nearly impossible to achieve. With little computational support for $h_\text{opt}(A_i)$ and no theoretical support for $h(A_i) = D_i^\intercal V_i^{-1}$, one might just simplify the entire process by letting $V_i$ have an independent correlation structure altogether. Our simulation studies in Section \ref{sec:simulation} also provide corroborative evidence supporting the use of an independent correlation structure when estimating the first-order effects. 

Although the discussion centered around cluster randomized trials, the DR-GEE2 estimator can be used in other settings when estimation of ICCs is of interest such as in reliability and agreement studies. We focused our discussion on binary outcomes, but the approach can be adapted to other types of exponential family outcomes in a straightforward manner by modifying the link function and variance function for the likelihood in question. When outcomes within clusters are not equicorrelated, our ICC estimators marginalize out factors which contribute to the non-exchangeable structure and returns an estimate which can be construed as an ``average" correlation.

We also proposed a stochastic algorithm to obtain the solutions to GEE2s. This new algorithm substantially increased convergence rate and reduced the run-times. It is in particular useful in settings where either the number of clusters or the size of clusters is large. Accurate estimation of ICCs in general requires adequate number of clusters relative to the cluster size. When the cluster size is large relative to the number of clusters, the standard algorithm suffers from convergence issues. The stochastic algorithm alleviates this problem by performing the estimation on a subsample from each cluster. 

In the presence of informative missing data, the correlation among missingness indicators needs to properly accounted for to arrive at the consistent estimators for the association parameters. We assumed rMAR in the current work. Future research on further relaxing this assumption would be useful. 
\newpage
\section{Appendices}
\subsection{Proof of CAN for DR estimator}\label{appendixDR}
It suffices to show $\E[\widetilde{\Phi}_i^Y(\textbf{Z}_i^*, \textbf{X}_i,\textbf{R}_i, \bb_Y^*, \ba_Y^*,\bb_R, \ba_R, \bb_Y, \ba_Y)] = 0$ from Eq \ref{augipwgee2} whenever the OM or PS is correctly specified. \\ \\ 
\textbf{\underline{Case 1: OM is correctly specified}} \\
Under this case, we have $\overline{\pi}_{ij} = \pi_{ij}$ and $\overline{\rho}_{ijj'} = \rho_{ijj'}$, so we have that $\E[\overline{\pi}_{ij}|A_i] = \pi_i^*$ and $\E[\overline{\rho}_{ijj'}^\dagger|A_i] = \rho_i^*$. From this, it is easy to verify $\E[E_i'|\textbf{R}_i, \textbf{X}_i, \textbf{Z}_i, A_i] = \textbf{0}$ and $\E[\zeta_i] = \textbf{0}$. Hence,
\begin{align*}
\E[\widetilde{\Phi}_i^Y] &= \E[D_i^\intercal V_i^{-1}W_i^RE_i' + \zeta_i] \\
&= \E[\E[D_i^\intercal V_i^{-1}W_i^RE_i'|\textbf{R}_i, \textbf{X}_i, \textbf{Z}_i, A_i]] + \E[\zeta_i] \\
&=\E[D_i^\intercal V_i^{-1}W_i^R\E[E_i'|\textbf{R}_i, \textbf{X}_i, \textbf{Z}_i, A_i]] + \textbf{0} \\
&= E[D_i^\intercal V_i^{-1}W_i^R\cdot \textbf{0}] \\
&= \textbf{0}
\end{align*}
\textbf{\underline{Case 2: PS is correctly specified}} \\
Under this case, we have $\overline{\pi}_{ij}^R = \pi^R_{ij}$ and $\overline{\rho}^R_{ijj'} = \rho^R_{ijj'}$; together, this implies that $\E[W_i^R] =\textbf{I}$. First, using the fact that $E'_i + E''_i = E_i$, we may express
\begin{align*}
\widetilde{\Phi}_i^Y &= D_i^\intercal V_i^{-1}W^R_i E_i -  D_i^\intercal V_i^{-1}W^R_i E_i'' - D_i^\intercal V_i^{-1} E_i'' + D_i^\intercal V_i^{-1}W^R_i E_i'' + \zeta_i \\
&=\underbrace{D_i^\intercal V_i^{-1}W^R_i E_i}_{\mathbb{Q}_1} + \underbrace{D_i^\intercal (V_i^{-1}-V_i^{-1}W_i^R)E_i''}_{\mathbb{Q}_2} + \underbrace{\zeta_i - D_i^\intercal V_i^{-1} E_i''}_{\mathbb{Q}_3}
\end{align*}
It now suffices to show $\E[\mathbb{Q}_1], \E[\mathbb{Q}_2], \E[\mathbb{Q}_3] = \textbf{0}$. We have $\E[\mathbb{Q}_1] = \textbf{0}$ by standard IPW-GEE2. Next,
\begin{align*}
\E[\mathbb{Q}_2] = \E[D_i^\intercal V_i^{-1}\E[\textbf{I}-W_i^R|\textbf{X}_i, \textbf{Z}_i^*]E_i''] = \E[D_i^\intercal V_i^{-1}(\textbf{I}-\textbf{I})E_i''] = \textbf{0}
\end{align*}
Finally,
\begin{align*}
\E[\mathbb{Q}_3] &= \E[\zeta_i] - \E[D_i^\intercal V_i^{-1} E_i'']\\
&= \E[\E[D_i^\intercal V_i^{-1} E_i''|\mathcal{D}_i\setminus A_i]] - \E[D_i^\intercal V_i^{-1} E_i''] \\
&= \E[D_i^\intercal V_i^{-1} E_i''] - \E[D_i^\intercal V_i^{-1} E_i''] \\
&= \textbf{0}
\end{align*}
Under certain regularity assumption defined in \cite{vaart2000}, we can
demonstrate with the Slutsky's theorem and the central limit theorem that any estimator
solving this Doubly Robust estimating equation is CAN.

\subsection{Pseudocode for Stochastic Algorithms}\label{appendixalgo}
\begin{algorithm}[H]
	\caption{S-IPW-GEE2 algorithm}
	\label{alg:IPWSGEE2}
	\begin{algorithmic}[1]
		\REQUIRE $\textbf{Y}, A_i, \textbf{Z}_i, \textbf{X}, \textbf{W}^R, \pi_S, \bg, \Omega$
		\STATE{$\bb_0, \ba_0 \gets \textbf{0}$}
		\FOR{$\omega = 0:(\Omega-1)$}
		\STATE{$U_i^\text{obs}\gets $ indices of observed $\textbf{Y}_i$ } for $i = 1:I$
		\STATE{$\upsilon_i \gets \lceil \pi_S |U_i^\text{obs}|\rceil$ } for $i = 1:I$
		\STATE{$s_i \sim \text{SRSWOR}(U_i^\text{obs}, \upsilon_i)$} for $i = 1:I$
		\STATE{$\widetilde{W}_{\beta i (\omega)}^R \gets \frac{m_i}{\upsilon_i}W_{\beta i (\omega)}^R [s_i]$} for $i = 1:I$
		\STATE{$\widetilde{W}_{\alpha i (\omega)}^R \gets \frac{m_i(m_i-1)}{\upsilon_i(\upsilon_i-1)}W_{\alpha i (\omega)}^R[(s_i)_2]$} for $i = 1:I$
		\STATE{$\widetilde{H}_{\beta i (\omega)} \gets \sum_{i=1}^{I}D_{\beta i(\omega)}^\intercal V_{\beta i(\omega)}^{-1}\widetilde{W}_{\beta i(\omega)}^{R}D_{\beta i(\omega)}$}
		\STATE{$\widetilde{G}_{\beta i (\omega)} \gets \sum_{i=1}^{I}D_{\beta i(\omega)}^\intercal V_{\beta i(\omega)}^{-1}\widetilde{W}_{\beta i(\omega)}^{R}E_{\beta i(\omega)}$}
		\STATE{$\widetilde{H}_{\alpha i (\omega)} \gets \sum_{i=1}^{I}D_{\alpha i(\omega)}^\intercal \widetilde{W}_{\alpha i(\omega)}^{R}D_{\alpha i(\omega)}$}
		\STATE{$\widetilde{G}_{\alpha i (\omega)} \gets \sum_{i=1}^{I}D_{\alpha i(\omega)}^\intercal \widetilde{W}_{\alpha i(\omega)}^{R}E_{\alpha i(\omega)}$}
		\STATE{$\bb_{(\omega+1)} \gets \bb_{(\omega)} + \gamma_\omega\widetilde{H}_{\beta i (\omega)}^{-1}\widetilde{G}_{\beta i (\omega)}$}
		\STATE{$\ba_{(\omega+1)} \gets \ba_{(\omega)} + \gamma_\omega\widetilde{H}_{\alpha i (\omega)}^{-1}\widetilde{G}_{\alpha i (\omega)}$}
		\ENDFOR
		\RETURN $\bb_{(\Omega)}, \ba_{(\Omega)}$
	\end{algorithmic}
\end{algorithm}	
\begin{algorithm}[H]
	\caption{S-DR-GEE2 algorithm}
	\label{alg:DRSGEE2}
	\begin{algorithmic}[1]
		\REQUIRE $\textbf{Y}, A_i, \textbf{Z}_i, \textbf{X}, \textbf{W}^R, \bpi, \br^\dagger, \pi_S, \bg, \Omega$
		\STATE{$\bb_0, \ba_0 \gets \textbf{0}$}
		\FOR{$\omega = 0:(\Omega-1)$}
		\STATE{$U_i^\text{obs}\gets $ indices of observed $\textbf{Y}_i$ } for $i = 1:I$
		\STATE{$U_i\gets $ indices of all $\textbf{Y}_i$ } for $i = 1:I$
		\STATE{$\upsilon_i \gets \lceil \pi_S |U_i^\text{obs}|\rceil$ } for $i = 1:I$
		\STATE{$\upsilon_i' \gets \lceil \pi_S |U_i|\rceil$ } for $i = 1:I$
		\STATE{$s_i \sim \text{SRSWOR}(U_i^\text{obs}, \upsilon_i)$} for $i = 1:I$
		\STATE{$s_i' \sim \text{SRSWOR}(U_i, \upsilon_i')$} for $i = 1:I$
		\STATE{$\widetilde{W}_{\beta i (\omega)}^R \gets \frac{m_i}{\upsilon_i}W_{\beta i (\omega)}^R [s_i]$} for $i = 1:I$
		\STATE{$\widetilde{W}_{\alpha i (\omega)}^R \gets \frac{m_i(m_i-1)}{\upsilon_i(\upsilon_i-1)}W_{\alpha i (\omega)}^R[(s_i)_2]$} for $i = 1:I$
		\STATE{$\widetilde{W}_{\beta i (\omega)}^{R'} \gets \frac{n_i}{\upsilon_i'}[s_i']$} for $i = 1:I$
		\STATE{$\widetilde{W}_{\alpha i (\omega)}^{R'} \gets \frac{n_i(n_i-1)}{\upsilon_i'(\upsilon_i'-1)}[(s_i')_2]$} for $i = 1:I$
		\STATE{$\widetilde{\zeta}_{\beta i (\omega)} \gets \sum_{a=0}^{1}p^a(1-p)^{1-a} D_{\beta i(\omega)}^\intercal (A = a)V_{\beta i(\omega)}^{-1}\widetilde{W}_{\beta i(\omega)}^{R'}E_{\beta i(\omega)}''(A = a)$}  for $i = 1:I$
		\STATE{$\widetilde{\zeta}_{\alpha i (\omega)} \gets \sum_{a=0}^{1}p^a(1-p)^{1-a} D_{\alpha i(\omega)}^\intercal (A = a)\widetilde{W}_{\alpha i(\omega)}^{R'}E_{\alpha i(\omega)}''(A = a)$}  for $i = 1:I$
		\STATE{$\widetilde{H}_{\beta i (\omega)} \gets \sum_{i=1}^{I}\sum_{a=0}^{1}p^a(1-p)^{1-a}D_{\beta i(\omega)}^\intercal(A=a) V_{\beta i(\omega)}^{-1}\widetilde{W}_{\beta i(\omega)}^{R}D_{\beta i(\omega)}(A=a)$}
		\STATE{$\widetilde{G}_{\beta i (\omega)} \gets \sum_{i=1}^{I}[D^\intercal_{\beta i(\omega)}V_{\beta i (\omega)}^{-1}\widetilde{W}_{\beta i (\omega)}^{R}E_{\beta i(\omega)}' + \widetilde{\zeta}_{\beta i(\omega)}]$}
		\STATE{$\widetilde{H}_{\alpha i (\omega)} \gets \sum_{i=1}^{I}\sum_{a=0}^{1}p^a(1-p)^{1-a}D_{\alpha i(\omega)}^\intercal(A=a) \widetilde{W}_{\alpha i(\omega)}^{R}D_{\alpha i(\omega)}(A=a)$}
		\STATE{$\widetilde{G}_{\alpha i (\omega)} \gets \sum_{i=1}^{I}[D^\intercal_{\alpha i(\omega)}\widetilde{W}_{\alpha i (\omega)}^{R}E_{\alpha i(\omega)}' + \widetilde{\zeta}_{\alpha i(\omega)}]$}
		\STATE{$\bb_{(\omega+1)} \gets \bb_{(\omega)} + \gamma_\omega\widetilde{H}_{\beta i (\omega)}^{-1}\widetilde{G}_{\beta i (\omega)}$}
		\STATE{$\ba_{(\omega+1)} \gets \ba_{(\omega)} + \gamma_\omega\widetilde{H}_{\alpha i (\omega)}^{-1}\widetilde{G}_{\alpha i (\omega)}$}
		\ENDFOR
		\RETURN $\bb_{(\Omega)}, \ba_{(\Omega)}$
	\end{algorithmic}
\end{algorithm}
\begin{algorithm}[H]
	\caption{DR-ParSGEE2 algorithm}
	\label{alg:ParSGEE2}
	\begin{algorithmic}[1]
		\REQUIRE $\textbf{Y}, \textbf{Z}^*, \textbf{X}, \textbf{W}^R, \bpi, \br^\dagger, \pi_S, \bg, \Omega, K$
		\FOR{$k = 1:K$}
		\STATE{$(\bb^{(k)}, \ba^{(k)})\gets$ DR-SGEE2($\textbf{Y}, \textbf{Z}^*, \textbf{X}, \textbf{W}^R, \bpi, \br^\dagger, \pi_S, \bg, \Omega$)}
		\ENDFOR
		\RETURN $\bb = \frac{1}{K}\sum_{k=1}^{K}\bb^{(k)}, \ba = \frac{1}{K}\sum_{k=1}^{K}\ba^{(k)}$
	\end{algorithmic}
\end{algorithm}
\newpage
\subsection{Time Complexity Proofs}\label{timecomplexproofs}
In proving the time-complexities associated with iterations of standard Fisher scoring or stochastic Fisher scoring, we make many uses of the following facts:\\ \\
\textbf{Fact 1:} The time complexity of multiplying matrix $A_{n\times m}$ and $B_{m\times p}$ is $\mathcal{O}(nmp)$. \\	\textbf{Fact 2:} The complexity of inverting an $n \times n$ matrix is $\mathcal{O}(n^3)$.\\
\textbf{Fact 3:} $\mathcal{O}(f(n)) + \mathcal{O}(g(n)) = \mathcal{O}(\max(f, g)(n))$. \\ \\
Omit the $R$ and $Y$ indices, for the computational complexity results are the same in both cases. Let $d_\beta = \dim(\beta), d_\alpha = \dim(\alpha)$. We make the assumptions that $d_\beta, d_\alpha, I$ are fixed; hence $\mathcal{O}(d_\beta) = \mathcal{O}(d_\alpha) = \mathcal{O}(I) = \mathcal{O}(1)$. Furthermore, we conduct the proofs as if we have no natural missingness in data, for proofs with the latter return the same complexities. We can decompose a covariance matrix $V = U^{1/2}CU^{1/2}$, where $C$ is a correlation matrix, and $U$ is a diagonal matrix with variance entries.

Table \ref{timecomplex} contains a total of 12 complexities. We break them down into four sub-theorems. Additionally, we require the assumption that $\pi_S \sim (\max_i n_i)^{-1}$; that is, our subsample size does not grow with respect to $n_i$.
\subsubsection*{Sub-theorem 1}\label{comp1} In the presence of standard Fisher scoring, an iteration of the GEE1 portion with
\begin{enumerate}
	\item[(i)] Arbitrary correlation matrix
	\item[(ii)] Equicorrelation matrix
	\item[(iii)] No correlation
\end{enumerate}
are of complexities $\mathcal{O}(\max_i n_i^3), \mathcal{O}(\max_i n_i), \mathcal{O}(\max_i n_i)$ respectively. \\ \\
\textit{Proof.} (i) Let us list the steps required in the computation:
\begin{enumerate}
	\item Computing $V_{\beta i\omega}^{-1}$:
	\begin{enumerate}
		\item Compute $C_{\beta i\omega}^{-1}$ and $U_{\beta i\omega}^{-1/2}$, which are of complexities $\mathcal{O}(n_i^3)$ and $\mathcal{O}(n_i)$, since $U_{\beta i\omega}$ is diagonal. The time complexity in computing $C_{\beta i\omega}^{-1}$, through either Gauss-Jordan elimination or Cholesky decomposition, is $\mathcal{O}(n_i^3)$ and cannot be sped up except through highly specialized numerically-optimized matrix algorithms (i.e. Coppersmith--Winograd algorithm).
		\item Compute $C_{\beta i\omega}^{-1}U_{\beta i\omega}^{-1/2}$. Because $U_{\beta i\omega}^{1/2}$ is diagonal, this becomes just multiplying the diagonal of $U_{\beta i\omega}^{-1/2}$ against each row of $C_{\beta i\omega}^{-1}$, and has complexity $\mathcal{O}(n_i^2)$.
		\item Left-multiply $C_{\beta i\omega}^{-1}U_{\beta i\omega}^{-1/2}$ with $U_{\beta i\omega}^{-1/2}$. This is also $\mathcal{O}(n_i^2)$.
	\end{enumerate}
	Hence, computing $V_{\beta i\omega}^{-1}$ has complexity $\mathcal{O}(n_i^3)$.
	\item Computing $H_{\beta i \omega}^{-1}$, having already computed $V_{\beta i\omega}^{-1}$:
	\begin{enumerate}
		\item Compute $V_{\beta i\omega}^{-1}D_{\beta i\omega}$. This has complexity $\mathcal{O}(d_\beta n_i^2) = \mathcal{O}(n_i^2)$. 
		\item Left-multiply $V_{\beta i\omega}^{-1}D_{\beta i\omega}$ by $D_{\beta i\omega}^\intercal$; this has complexity $\mathcal{O}(d_\beta^2n_i) = \mathcal{O}(n_i)$.
		\item Invert the resulting $D_{\beta i\omega}^\intercal V_{\beta i\omega}^{-1}D_{\beta i\omega}$. This is time complexity $\mathcal{O}(d_\beta^3) = \mathcal{O}(1)$.
	\end{enumerate}
	Hence, complexity in computing $H_{\beta i \omega}$ is $\mathcal{O}(n_i^2)$.
	\item Computing $G_{\beta i \omega}$, having already computed $V_{\beta i\omega}^{-1}$:
	\begin{enumerate}
		\item All steps are almost the same as computing $H_{\beta i \omega}$, except for 2(a), where we have  $V_{\beta i\omega}^{-1}E_{\beta i\omega}$, which is still $\mathcal{O}(n_i^2)$
	\end{enumerate}
	Overall, computing $G_{\beta i \omega}$ is $\mathcal{O}(n_i^2)$
	\item  Computing $H_{\beta i \omega}^{-1}G_{\beta i \omega}$, having already computed $H_{\beta i \omega}^{-1}$ and $G_{\beta i \omega}$, is just $\mathcal{O}(d_\beta) = \mathcal{O}(1)$.
\end{enumerate}
Overall, steps 1 -- 4 is of $\mathcal{O}(n_i^3)$, due to computing $V_{\beta i \omega}^{-1}$.
\begin{enumerate}[resume]
	\item Perform steps 1 -- 4 for each $i$. The time complexity is $\sum_{i=1}^{I}\mathcal{O}(n_i^3) = \mathcal{O}(\max_i n_i^3)$.
	\item Summing up $H_{\beta i \omega}^{-1}G_{\beta i \omega}$ is $\mathcal{O}(I) = \mathcal{O}(1)$, and then adding this resulting quantity is $\mathcal{O}(1)$.
\end{enumerate}
Overall, we have $\mathcal{O}(\max_i n_i^3)$. \\ \\
(ii) Since $C_{\beta i\omega}$ is equicorrelated, we have that
\begin{align*}
C_{\beta i\omega}^{-1} = (1-\rho_i)^{-1}\left(\mathbf{I}_{n_i} - \frac{\rho_i}{1 + (n-1)\rho_i}J_{n_i}\right)
\end{align*}
by Woodbury's formula, where $J_{n_i}$ is an $n_i \times n_i$ matrix of 1's. Hence, in computing $H_{\beta i \omega} = D_{\beta i\omega}^\intercal V_{\beta i\omega}^{-1}D_{\beta i\omega}$, we would compute
\begin{align*}
\underbrace{(1-\rho_i)^{-1}D_{\beta i\omega}^\intercal U_{\beta i\omega}^{-1} D_{\beta i\omega}}_{Q_1} - \underbrace{\frac{\rho_i}{(1 + (n_i-1)\rho_i)(1-\rho_i)} D_{\beta i\omega}^\intercal U_{\beta i\omega}^{-1/2}J_{n_i}U_{\beta i\omega}^{-1/2}D_{\beta i\omega}}_{Q_2}
\end{align*}
Since $U_{\beta i\omega}^{-1}$ is diagonal, we can perform an element-wise product with the diagonal, and hence computation of $Q_1$ is $\mathcal{O}(n_i)$. In computing $Q_2$, notice that to compute $J_{n_i}U_{\beta i\omega}^{-1/2}D_{\beta i\omega}$ is to 
\begin{enumerate}
	\item Perform $U_{\beta i\omega}^{-1/2}D_{\beta i\omega}$, which can be done through element-wise product.
	\item Sum each column of the resulting $U_{\beta i\omega}^{-1/2}D_{\beta i\omega}$ into a row vector.
	\item Repeat each row $n_i$ times into a matrix.
\end{enumerate}
This has time complexity $\mathcal{O}(n_i)$. Then, left-multiplying this quantity by $U_{\beta i\omega}^{-1/2}$ and then again by $D_{\beta i\omega}^\intercal$ is $\mathcal{O}(n_i)$ and $\mathcal{O}(d^2_\beta n_i) = \mathcal{O}(n_i)$. Overall, computing $H^{-1}_{\beta i \omega}$ is now $\mathcal{O}(n_i)$. Analogous steps can be done to calculate $G_{\beta i \omega}$, which is now $\mathcal{O}(n_i)$. The rest of the proof follows steps 4 -- 6 of (i), which results in $\mathcal{O}(\max_i n_i)$. \\ \\
(iii) For no correlation, inverting $V_{\beta i \omega}$ requires inverting the diagonal entries; this is still of complexity $\mathcal{O}(n_i)$. Rest of the proof follows as (i). \hfill $\boxed{}$ 
\subsubsection{Sub-theorem 2} \label{comp2} In the presence of standard Fisher scoring, an iteration of the GEE2 portion with
\begin{enumerate}
	\item[(i)] Arbitrary correlation matrix
	\item[(ii)] Equicorrelation matrix
	\item[(iii)] No correlation
\end{enumerate}
are of complexities $\mathcal{O}(\max_i n_i^6)$, $\mathcal{O}(\max_i n_i^2)$, $\mathcal{O}(\max_i n_i^2)$ respectively. \\ \\
\textit{Proof.} All rows and columns in the proofs for GEE1 now have lengths $\binom{n_i}{2} \sim n_i^2$ in place of $n_i$. Hence, all exponents in computational complexities in Theorem \ref{comp1} are doubled. \hfill $\boxed{}$ \\ \\	
Now, let's continue with stochastic Fisher scoring. Define $D_{\beta i\omega}^\text{sub}, E_{\beta i\omega}^\text{sub}$ as the resulting $D_{\beta i\omega}, E_{\beta i\omega}$ with only rows corresponding to subsample $s_i$; we see that, the dimensions of these matrices are now $\upsilon_i \times d_\beta$ and $\upsilon_i \times 1$, respectively. Let $\widetilde{W}_{\beta i (\omega)}^{R\text{sub}}$ equal $\widetilde{W}_{\beta i (\omega)}^{R}$ except with both rows and columns associated with zero diagonal elements removed; this has dimension $\upsilon_i \times \upsilon_i$. We can analogously define this for $D_{\alpha i\omega}^\text{sub}, E_{\alpha i\omega}^\text{sub}, \widetilde{W}_{\alpha i\omega}^{R\text{sub}}$, where any dimension with a $\binom{n_i}{2}$ is replaced with $\binom{\upsilon_i}{2}$. 
\subsubsection{Sub-theorem 3} \label{comp3} In the presence of stochastic Fisher scoring, an iteration of the GEE1 portion with
\begin{enumerate}
	\item[(i)] Arbitrary correlation matrix
	\item[(ii)] Equicorrelation matrix
	\item[(iii)] No correlation
\end{enumerate}
will be of complexities $\mathcal{O}(\max_i n_i^3), \mathcal{O}(\max_i n_i), \mathcal{O}(1)$ respectively. \\ \\
\textit{Proof.} (i) We cannot exploit sparsity here, for the largest complexity object, $V^{-1}_{\beta i\omega}$, would still need to be computed, which is $\mathcal{O}(n_i^3)$. \\ \\
(ii) Let's list again the steps in computing the quantities.
\begin{enumerate}
	\item Computing $\widetilde{H}_{\beta i \omega}^{-1}$:
	Using Woodbury's formula, the computation of $\widetilde{H}_{\beta i \omega}$ would be
	\begin{align*}
	(1-\rho_i)^{-1}D_{\beta i\omega}^\intercal U_{\beta i\omega}^{-1} \widetilde{W}_{\beta i\omega}^{R} D_{\beta i\omega} - \frac{\rho_i}{(1 + (n_i-1)\rho_i)(1-\rho_i)} D_{\beta i\omega}^\intercal U_{\beta i\omega}^{-1/2}J_{n_i}U_{\beta i\omega}^{-1/2} \widetilde{W}_{\beta i\omega}^{R} D_{\beta i\omega}
	\end{align*}
	Exploiting sparsity, this is the same as
	\begin{align*}
	&\underbrace{(1-\rho_i)^{-1}D_{\beta i\omega}^\intercal (U_{\beta i\omega}^\text{sub})^{-1}\widetilde{W}_{\beta i\omega}^{R\text{sub}} D_{\beta i\omega}^\text{sub}}_{\widetilde{Q}_1} -\underbrace{\frac{\rho_i}{(1 + (n_i-1)\rho_i)(1-\rho_i)} D_{\beta i\omega}^\intercal U_{\beta i\omega}^{-1/2} J_{n_i \times \upsilon_i}(U_{\beta i\omega}^\text{sub})^{-1/2} \widetilde{W}_{\beta i\omega}^{R\text{sub}} D_{\beta i\omega}^\text{sub}}_{\widetilde{Q}_2}
	\end{align*}
	\begin{enumerate}
		\item Computing $\widetilde{Q}_1$ first performs the following steps:
		\begin{align*}
		\widetilde{W}_{\beta i\omega}^{R\text{sub}} D_{\beta i\omega}^\text{sub} \mapsto  U_{\beta i\omega}^{-1} \widetilde{W}_{\beta i\omega}^{S} D_{\beta i\omega} \mapsto D_{\beta i\omega}^\intercal U_{\beta i\omega}^{-1} \widetilde{W}_{\beta i\omega}^{S} D_{\beta i\omega} \mapsto (1-\rho_i)^{-1}D_{\beta i\omega}^\intercal U_{\beta i\omega}^{-1} \widetilde{W}_{\beta i\omega}^{S} D_{\beta i\omega}
		\end{align*}
		which sequentially, conditioned on performing the previous computation, is $\mathcal{O}(d_\beta \upsilon_i)$, $\mathcal{O}(d_\beta \upsilon_i)$, $\mathcal{O}(d_\beta^2 \upsilon_i)$, and $\mathcal{O}(d_\beta^2)$. The sum of these three complexities is $\mathcal{O}(\upsilon_i)$.
		\item Computing $Q_2$ first performs the following steps:
		\begin{align*}
		\widetilde{W}_{\beta i\omega}^{R\text{sub}} D_{\beta i\omega}^\text{sub} &\mapsto (U_{\beta i\omega}^\text{sub})^{-1/2} \widetilde{W}_{\beta i\omega}^{R\text{sub}} D_{\beta i\omega}^\text{sub} \\
		&\mapsto J_{n_i \times \upsilon_i}(U_{\beta i\omega}^\text{sub})^{-1/2} \widetilde{W}_{\beta i\omega}^{R\text{sub}} D_{\beta i\omega}^\text{sub} \\
		&\mapsto U_{\beta i\omega}^{-1/2} J_{n_i \times \upsilon_i}(U_{\beta i\omega}^\text{sub})^{-1/2} \widetilde{W}_{\beta i\omega}^{R\text{sub}} D_{\beta i\omega}^\text{sub} \\
		&\mapsto D_{\beta i\omega}^\intercal U_{\beta i\omega}^{-1/2} J_{n_i \times \upsilon_i}(U_{\beta i\omega}^\text{sub})^{-1/2} \widetilde{W}_{\beta i\omega}^{R\text{sub}} D_{\beta i\omega}^\text{sub} \\
		&\mapsto \frac{\rho_i}{(1 + (n_i-1)\rho_i)(1-\rho_i)} D_{\beta i\omega}^\intercal U_{\beta i\omega}^{-1/2} J_{n_i \times \upsilon_i}(U_{\beta i\omega}^\text{sub})^{-1/2} \widetilde{W}_{\beta i\omega}^{R\text{sub}} D_{\beta i\omega}^\text{sub}
		\end{align*}
		The time complexities of each step is $\mathcal{O}(d_\beta \upsilon_i)$, $\mathcal{O}(d_\beta \upsilon_i)$, $\mathcal{O}(d_\beta \upsilon_i)$, $\mathcal{O}(d_\beta n_i)$, $\mathcal{O}(d_\beta^2 n_i)$, and $\mathcal{O}(d_\beta^2)$. Notice that the third step cannot be simplified due to the $J_{n_i \times \upsilon_i}$ matrix separating $D_{\beta i\omega}^\intercal$ and $\widetilde{W}_{\beta i\omega}^{R\text{sub}}$. 
		\item Inverting $H_{\beta i \omega}$ is again $\mathcal{O}(d_\beta^3)$, which is dominated by the other steps.
	\end{enumerate}
	Hence, calculating $H_{\beta i \omega}^{-1}$ is $\mathcal{O}(n_i)$.
	\item Steps in computing $G_{\beta i \omega}^{-1}$ are analogous to step 1, and also $\mathcal{O}(n_i)$
\end{enumerate}
Repeat steps 4 -- 6 of Theorem \ref{comp1} (i), we again have $\mathcal{O}(\max_i n_i)$. \\
\textbf{Remark:} For the cases of a general or equicorrelated $C_{\beta i\omega}$, the time complexities of standard and stochastic Fisher scorings are the same. Intuitively, although we want to feed a subset of the data into the scoring equations, we cannot make full use of sparsity because the inverse-covariance matrix $V_{\beta i \omega}^{-1}$ forces a ``mixing" of all the observations, including into missing vector slots. The next two settings no longer have any correlations, and hence we can make full use of sparsity. \\ \\
(iii) We present just the proof of computing $\widetilde{H}_{\beta i \omega}$, since this and $\widetilde{G}_{\beta i \omega}$ are bottlenecks in the computation, and both have the same complexities. We now just need to compute
\begin{align*}
D_{\beta i\omega}^\intercal U_{\beta i\omega}\widetilde{W}_{\beta i\omega}^{R}D_{\beta i\omega} = (D_{\beta i\omega}^\text{sub})^\intercal U_{\beta i\omega}^\text{sub}\widetilde{W}_{\beta i\omega}^{R\text{sub}}D_{\beta i\omega}^\text{sub}
\end{align*}
Sequentially, the steps in computing 
\begin{align*}
\widetilde{W}_{\beta i\omega}^{R\text{sub}}D_{\beta i\omega}^\text{sub} \mapsto U_{\beta i\omega}^\text{sub}\widetilde{W}_{\beta i\omega}^{R\text{sub}}D_{\beta i\omega}^\text{sub} \mapsto (D_{\beta i\omega}^\text{sub})^\intercal U_{\beta i\omega}^\text{sub}\widetilde{W}_{\beta i\omega}^{R\text{sub}}D_{\beta i\omega}^\text{sub}
\end{align*}
are of $\mathcal{O}(d_\beta \upsilon_i), \mathcal{O}(d_\beta\upsilon_i), \mathcal{O}(d_\beta^2\upsilon_i)$; overally, this is of time complexity $\mathcal{O}(\upsilon_i) = \mathcal{O}(1)$, if we choose $\pi_S \sim (\max_i n_i)^{-1}$. \hfill $\boxed{}$ 
\subsubsection{Sub-theorem 4}  \label{comp4} In the presence of stochastic Fisher scoring, an iteration of the GEE2 portion with
\begin{enumerate}
	\item[(i)] Arbitrary correlation matrix
	\item[(ii)] Equicorrelation matrix
	\item[(iii)] No correlation
\end{enumerate}
will be of complexities $\mathcal{O}(\max_i n_i^6), \mathcal{O}(\max_i n_i^2), \mathcal{O}(1)$ respectively. \\ \\
\textit{Proof.} Apply Sub-theorem 3 with $\upsilon_i$ replaced with $\binom{\upsilon_i}{2} \sim \upsilon_i^2$, and we are done. \hfill $\boxed{}$ \newpage
\bibliography{ICC}
\bibliographystyle{asa}
\end{document}